\documentclass[journal]{IEEEtran}
\usepackage{ifpdf}

\usepackage{cite}

\ifCLASSINFOpdf
\usepackage[pdftex]{graphicx}
\graphicspath{{../figs/}}
\DeclareGraphicsExtensions{.pdf,.jpeg,.png,.eps}
\else
\usepackage[dvips]{graphicx}
\graphicspath{{../figs/}}
\DeclareGraphicsExtensions{.eps}
\fi

\ifCLASSOPTIONcompsoc
 \usepackage[caption=false,font=normalsize,labelfont=sf,textfont=sf]{subfig}
\else
 \usepackage[caption=false,font=footnotesize]{subfig}
\fi

\usepackage{dblfloatfix}

\ifCLASSOPTIONcaptionsoff
\usepackage[nomarkers]{endfloat}
\let\MYoriglatexcaption\caption
 \renewcommand{\caption}[2][\relax]{\MYoriglatexcaption[#2]{#2}}
\fi

\usepackage{array}

\usepackage{amsthm}
\usepackage{amsmath}
\usepackage{amsfonts}
\usepackage{amssymb}
\usepackage{bm}

\usepackage{booktabs}
\usepackage{threeparttable}
\usepackage{multirow}
\usepackage{csquotes}

\newtheorem{thm}{Theorem}
\newtheorem{defn}{Definition}

\theoremstyle{remark}
\newtheorem{rem}{Remark}

\newtheorem{assum}{Assumption}
\renewcommand{\IEEEQED}{\IEEEQEDopen}

\usepackage{algorithm}
\usepackage{algorithmicx}
\usepackage{algpseudocode}
\usepackage{xfrac}
\usepackage{url}

\usepackage{cleveref}

\begin{document}

\title{\enquote{Closed Proportional-Integral-Derivative-Loop Model} Following Control}
%
%
%

\author{Oluwasegun~A.~Somefun~\IEEEmembership{Member,~IEEE,}
        Kayode~F.~Akingbade
        and~Folasade~M.~Dahunsi
\thanks{This research was not funded.}%
\thanks{O. A. Somefun is with the Department
of Computer Engineering,  Federal University of Technology, Akure,
Ondo, Nigeria, P.M.B 704. e-mail: (oasomefun@futa.edu.ng).}
\thanks{F.M. Dahunsi is with the Department
of Computer Engineering,  Federal University of Technology, Akure,
Ondo, Nigeria, P.M.B 704. e-mail: (fmdahunsi@futa.edu.ng).}
\thanks{K. F. Akingabde is with the Department
of Electrical and Electronics Engineering,  Federal University of Technology, Akure, P.M.B 704. e-mail: (kfakingbade@futa.edu.ng).}%
\thanks{Manuscript received April X, 2020; revised X.X.,2020. Corresponding Author: Oluwasegun Somefun}
}


\markboth{}%
{Somefun \MakeLowercase{\textit{et al.}}:\enquote{Closed Proportional-Integral-Derivative-Loop Model}
Following Control}


\maketitle
\begin{abstract}
The proportional-integral-derivative (PID) control law is often overlooked as a computational imitation of the critic control in human decision. This paper provides a formulation to remedy this problem. Further, based on the characteristic settling-behaviour of dynamical systems, the \enquote{closed PID-loop model} following control (CPLMFC) method is introduced for automatic PID design. Also, a method for closed-loop settling-time identification is provided. The CPLMFC algorithm and some recommended guidelines are given for setting the critic weights of the PID. Finally, two representative case-studies are simulated. Both the theoretical results and simulation results (via performance indices) illustrate that the CPLMFC can guarantee both accurate and stable closed-loop adaptive PID control performance in real-time.
\end{abstract}

%
%


\IEEEpeerreviewmaketitle

\section{Introduction}\label{sec1}
The need for guaranteed stable and accurate closed-loop performance, especially in motor-control tasks becomes much more evident in times of a threatening pandemic like COVID-19. For instance, a lot of dynamical systems such as the same models of mobile robots, manipulators, and aerial robots, need to be deployed to repetitively do the same tasks. There needs to be a guarantee that they will exhibit the same desired motor response. In this case, a well-designed feedback controller then becomes a real money-savings and life-savings investment against increased risk and spread of infections \cite{yangCoronavirusPandemicCall2020}. In practise, a lot of these automation tasks still fundamentally rely on a form of  Proportional-Integral-Derivative (PID) control. PID control is a common, yet powerful representation of the universal feedback control law \cite{vagia_pid_2012, silva_pid_2005}. Research on PID control design has therefore attracted widespread theoretical and practical interest \cite{vilanovaPIDControlThird2012, astromAdvancedPIDControl2018, wang_approach_2018}. However, the best control design approach to tuning a closed PID-loop in order to achieve its best performance in terms of stability and accuracy remains an open question \cite{leiFeedbackUncertaintyBasic2019}. Closed-loop performance in terms of accuracy and stability are therefore critical to the performance of a whole dynamical system setup \cite{sung_process_2009, ang_pid_2005}. Accordingly, this paper introduces the \enquote{Closed PID-Loop Model} Following Control (CPLMFC) method for real-time adaptive computer control of stabilizable dynamical (physical) systems.

Consequently, we argue, in this paper, that by its structure, the PID law implicitly has an ideal (model) closed-loop response that it tries to reference. Therefore, the PID can in real-time automatically tune itself using this reference model called the closed-PID loop model (CPLM). This model is implicit and is a model case of dominant eigenvalue assignment of complex-conjugate poles. We show that the key to designing this model is its natural-frequency. We make a connection between this frequency parameter and the identified closed-loop settling-time. This is the closed-loop settling-time of the dynamical system under control inside a closed PID-loop. This settling-time identification is significant, because with a roughly accurate estimation, the dominant eigenvalues of the closed PID-loop system can be assigned to stably and accurately follow the CPLM.

Specifically, for practical physical systems with finite settling behaviour (FSB), an implication of this is that manufacturers can indicate the average estimated input-output settling-time in data-sheets to further streamline PID control for end-users.

Significantly, the CPLMFC method does not use the knowledge of a representative model of the dynamical system. Also it does not explicitly minimize an objective (cost) function. This contrasts with mainstream control methods, that have been applied to automatic PID control design, such as the general model predictive control (MPC) \cite{klaucoMPCBasedReferenceGovernors2019}, H-infinity \cite{diaz-rodriguezAnalyticalDesignPID2019}, and quantitative feedback theory (QFT)\cite{comasolivasAutomaticDesignRobust2012,mercaderRobustPIDDesign2017} methods.

On the whole, we show that: the CPLMFC method implicitly minimizes the perfect model following objective function of the CPLM. The CPLM is the ideal closed-PID loop response from the viewpoint of the PID control law. Therefore, emphasized next are the main contributions made in this paper.

\subsection{Main Contributions}\label{Smaincontrib}
To set this work in proper context, first a background of related literature is discussed in section~\ref{sec2}.
In section~\ref{sec3}, we start by formulating the PID control problem. Given a PID law that predicts a bounded input value, the central problem is how to design the three PID parameters, so that the moving single output of a dynamic system in the closed PID-loop approaches the bounded desired value.
\subsubsection{Critic PID}
An important aspect of intelligent decision or prediction is criticism \cite{marino_pid_2019,wangAdaptiveCriticControl2019}. In section~\ref{Scritic}, we consider the idea of the PID as a criticism-based two degree-of-freedom (2DOF) structure for the first-time in literature. A compact representation is also proved in section~\ref{Scomppid}. The PID formulation with critic weights imitates the basic critic functionality in the human approach to decision-making or learning. The introduced three critic weights are manual or automatic basis for criticising the three core PID control functions (that is the proportional, integral and derivative output terms).
\subsubsection{Closed PID-Loop Model Following Control}
In section~\ref{sec4}, we continue by considering a solution to the formulated problem from the view-point of model following (MF) and its connection to settling behaviour (SB). Note that the theoretical SB of physical systems may be classified as either finite (FSB) or infinite (ISB) \cite{raoNaiveControlDouble2001}. Obviously, the idea of designing a control system to follow (or mimic) the output of a desired model (termed the reference model) is not new in control theory \cite{okuyamaDiscreteControlSystems2014}. For example, modern model reference adaptive control concepts are special cases.

MF through a reference model is attractive because it sidesteps the difficulty of matching design specifications to controller parameters. A central notion in MF is perfect model following (PMF). It is a condition that if achieved, defines whether the physical system can follow the reference model exactly. However, due to physical limitations and allowable complexity of dynamic compensation, it is not always clear how this appropriate reference model should be chosen to satisfy the PMF \cite{gaoTechniquesReconfigurableControl1996}.

We argue that, if a closed-PID loop control system is stabilizable, this implies that the PID control law implicitly has a reference model that it follows. We start with some assumptions in section~\ref{Sassump}. In section~\ref{Scpla}, we then proceed to analyze the generic closed PID-loop transfer function. The main result is the identification of the nominal \enquote{closed PID-loop model} (CPLM) based on the loop-gain condition. The loop gain condition states that provided the loop-gain magnitude value of the closed-PID loop system can dominate one (1), the closed-loop system can be approximated as the nominal \enquote{closed PID-loop model}. We interpret this as the PID's idealized behavioural response of a closed PID-loop system.

Then, we continue, by showing that the PID is driven by this CPLM view to stabilize the closed loop system. Again, note that the nominal CPLM is independent of both the order and model parameters of its corresponding dynamical system (with model parameters assumed unknown under control). The PID (with one derivative term) attempts to have control over two desired pole locations and at most two desired zero locations in the \textit{s}-plane. Hence, it tries to approximate the closed PID-loop system to a second-order dominant dynamic system. The implication of this is that, the problem is implicitly reduced to a case of dominant eigenvalue assignment through the natural frequency, and damping of the nominal CPLM. Therefore, it is shown in section~\ref{Skikd} that the CPLM automatically gives settings for both the integral and derivative time constants to minimize the MF error for stability.

In section~\ref{Snormresp}, we continue by analyzing the normalized settling-behaviour of the nominal CPLM. The main results together with that in section~\ref{Swn} show that provided the dynamical system has a FSB, then settling-time can be used as a design parameter to set the natural frequency of the CPLM. In finite time, we want to stabilize the moving output of the physical system before it disintegrates. These factors motivate design of the closed-loop settling-time identification process discussed in section~\ref{Stsident}.

Further, in section~\ref{Sadaptkp}, we show using a state-space design methodology that the MF problem then reduces to that of finding a stabilizing proportional gain that minimises the MF error. We show the conditions under which PMF is possible. Then, we show that this optimal stabilizing proportional gain set can be asymptotically approached by an adaptive proportional gain update rule. However in cases, such as when delays become significant, the initial adaptive solution does not necessarily achieve PMF. The process of deriving settings for control parameters by correlating with representative processes is not new in controls. Since, a proportional gain in the stabilizing set may not be asymptotically reached with the initial adaptive solution, the study of a normalized first-order plus dead-time system approximation is used to show how the proportional gain set can be quickly reached. We use this approach to derive a rational function. This curve-fitted function is used as the bounding limit for a newly constructed nonlinear adaptive function that supports the initial adaptive proportional gain solution.

Furthermore, the CPLMFC algorithm in compact form for computer control is presented in section~\ref{Salgcplmfc}. Also, we include some recommended guidelines based on simulations for manually setting the critic weights for PID control.

The results in this paper provide a method for automatic robust PID design in the model following sense based on the characteristic settling behaviour of stabilizable dynamical systems. We have appropriately termed this the \enquote{Closed PID-Loop Model} Following Control (CPLMFC) method. The main idea is that if we can appropriately identify a stabilizable proportional gain for the closed PID-loop model (CPLM), and appropriate critic weights, then both stable and accurate settling behaviour can be guaranteed for the actual closed PID-loop system.

\subsubsection{Applications} In section~\ref{sec5}, we show that with critic PID control, the CPLMFC method can in real-time guarantee stable and accurate adaptive control computations of the three main PID parameters. We carry out numerical control simulations (via MATLAB) on two representative cases. For the first case: a delay-free, stable system, the benchmark example of a normalized third-order linear dynamical system is used. For the second case: a time-delayed integrating, non-linear and uncertain system in form of a real-life permanent-magnet linear motor model is used. Illustratively, through common performance indices: closed-loop maximum overshoot, settling-time ($1\%$) and integral error indices; we show that both accurate and stable closed-loop performance can be achieved using the CPLMFC method in such representative cases.

Finally, some discussions in section~\ref{secdiss}, and then this paper is concluded in section~\ref{secconc}.

\section{Related Works}\label{sec2}
In comparison with other formal control algorithms, the PID control law has been specially regarded as ubiquitous
\cite{samad_survey_2017}. A recurring motivation for this is because, although an apparently simple control strategy, PID control
has in application to many practical problems, consistently provided simple, least-cost, and satisfactory robust
control performance \cite{peretz_randomized_2018, bucz_advanced_2018,li_research_2016,yu_chapter_2018}.

In terms of representation, a straightforward and more general computational model of the PID control algorithm has a 2DOF control structure \cite{abdelaty_fixed_2018, viteckova_2dof_2015-1, wangNewDesignStrategy2012, astromPIDControllersTheory1995}.
This is the integer-valued number of closed-loop transfer functions that are present in the controller's structure and can be adjusted independently\cite{araki_two-degree--freedom_2003}. An implicit model following is achieved in some sense using this form \cite{astromPIDControllersTheory1995}.

On the other hand, the design problem (tuning) of the optimum set of PID parameters have been categorized as an NP-hard problem in terms of complexity \cite{koszakaIdeaUsingReinforcement2006}. This implies there is no unique solution. It also means that tuning for guaranteed accurate and stable control performance can be burdensome even for very common servomechanism applications \cite{roux-oliveira_extremum_2019,grimholt_optimal_2018,killingsworth_pid_2006}. Consequently, in \cite{astromAdvancedPIDControl2018}, two important problems were highlighted: One, the place of the PID algorithm as an active research in adaptive control. A concise review on adaptive control systems can be found in \cite{blackAdaptiveSystemsHistory2014}. Two, the need for automatic PID tuning algorithms instead of simple rules.

It is known that mainstream PID control design methods available in the control literature, are highly dependent on the knowledge of derived or fitted mathematical model approximations (plant models) of the actual physical system \cite{astrom_design_2017, odwyer_handbook_2009, astromAdvancedPIDControl2006, ellisControlSystemDesign2012}. The parameters of these plant models are then explicitly used as design constants to set the PID parameters \cite{li_research_2016,levine2011,sung_process_2009,visioli_practical_2006}. Also, corresponding adaptive designs, achieve parameter estimations using the derived plant model parameters \cite{dastjerdi_tuning_2018,jantzen_turning_2016-1}. From a dynamical systems perspective, this modeling approach is useful and more so, reasonable in some sense \cite{samad_survey_2017,samad_new_2013-1, bai_classical_2019,lynch_pid_2016,forrai_embedded_2013}. Notwithstanding, these models are imperfect representations of the physical systems, which although may operate in a linear sense, are actually both non-linear and uncertain. Therefore, these approaches may not fully exploit the best quality of control accuracy using the PID. According to a ASEA Brown Boveri (ABB) survey in \cite{koelschTuningToolsMaintain2014}, after sometime these tuning methods, which are commercially available become unreliable.

Deviating from model-based conventions are model-free and essentially data-driven control design methods. Among them, one popular technique is Relay Auto-tuning \cite{astromAdvancedPIDControl2018}. Relay Auto-tuning, uses modified Ziegler-Nichols (Z-N) tuning rules to automate the frequency response identification of a dynamical system. It relies on describing nonlinear function analysis, and so frequency limit cycles are common. The accuracy of its identified critical frequency point of oscillation is still a subject of active research \cite{zengResearchImprovedAutoTuning2019,hornseyReviewRelayAutotuning2012}. Also, notable in the model-free category, are design methods which explicitly minimize a cost function for real-time optimization, such as the unfalsified method, iterative feedback tuning (IFT) \cite{hoRelayAutotuningPID2003}, iterative learning control (ILC) \cite{mooreEditorialSpecialIssue2000}, and extremum-seeking method summarized in \cite{killingsworth_pid_2006}. For the IFT, multiple closed loop experiments have to be carried out. The ILC approach leverages repitition. To generally achieve perfect tracking performance, it requires physical systems to satisfy an identical initialization condition, which is a restrictive condition\cite{guanIterativeLearningControl2014}. Further, approaches that exploit the theory of fuzzy logic, neural network, and reinforcement learning for PID design exist \cite{shipmanReinforcementLearningDeep2019, mendelUncertainRuleBasedFuzzy2017}. Fuzzy-logic PID control techniques fuzzify the system error states. It is afflicted by time-consuming rule parameter optimization problems \cite{bai_classical_2019}.

Also, there are methods that mix both model-based and mode-free paradigms for PID control design. Proposed in \cite{jantzen_turning_2016-1}, and originating from fuzzy control study is a less known but promising mixed approach. The central argument in this approach is to simplify modeling by considering the physical meaningful performance specification of desired settling-time based on operator knowledge and system operating behaviour for tuning PIDs. Generally, in this mixed class, most approaches use either or both ideas of model following and adaptive control. Examples are direct (or indirect) model reference adaptive methods \cite{blackAdaptiveSystemsHistory2014} and existing self-tuning methods \cite{bobalDigitalSelftuningControllers2005, aggarwalSelfTuningAnalogProportionalIntegralDerivative2006}. Most of these methods usually suffer from poor transient performance or (and) involve some complexity of online parameter estimations.

Many recent works on PID control such as in
\cite{ekinciImprovedKidneyInspiredAlgorithm2019,hekimoglu_optimal_2019,mandava_optimal_2019,almabrok_fast_2018} have employed metaheuristic (evolutionary) optimization algorithms as new PID tuning methods. These methods consider different objective functions and performance constraints such as stability, bandwidth, and robustness. They usually still often require the identification of a plant model for the physical system. At the basic level, they can be viewed as stochastic search methods. Although promising, they have not found scalable use in the industry, since any serious uncertainty in the parameters of the system can lead to instability, and, in some cases, the converged solutions are less than optimal \cite{liCrossroadArtificialIntelligence2019, dastjerdi_tuning_2018}.

Dominant eigenvalue assignment, crudely known as pole-placement, is a class of general PID control design methods in the control theory literature, treated in works such as \cite{dattaStructureSynthesisPID2013,han_pid_2018,diaz-rodriguezAnalyticalDesignPID2019,keelRobustnessFragilityHigh2016}. Note that, this approach is attractive since performance specifications such as stability and dynamic performance directly correlate to the assignment and distribution of dominant closed-loop eigenvalues (poles). 

As highlighted in \cite{leiFeedbackUncertaintyBasic2019, chenControllerParameterOptimization2019, roux-oliveira_extremum_2019}, current understanding and rationale for the remarkable practical effectiveness and capability of PID control is still limited. Therefore further research needs to focus on dynamical systems as nonlinear and uncertain. As such, new unconventional approaches that automate tuning and streamline the model-free system identification approach for tuning in PID control would be of great advantage to control theory and practise, improve the PID's widespread adoption and contribute to better product quality \cite{chenControllerParameterOptimization2019, jantzen_turning_2016-1}.

\begin{figure}[!b]
\centering
\includegraphics[width=2.0in]{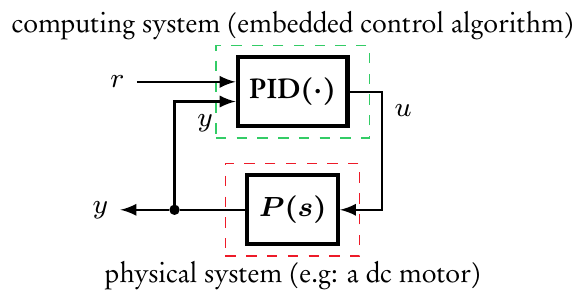}
\caption{Overview of a closed-PID control software loop.}
\label{fig_prob}
\end{figure}
\section{Problem Formulation}\label{sec3}
In this section, the control problem is formulated and a simple critic PID computational model is introduced.

Let us consider a single-input single-ouput (SISO) dynamical system with assumed unknown dynamical model parameters, which can be written in reduced form as (\ref{upltss1}).
\begin{IEEEeqnarray}{c}
\begin{cases}\label{upltss1}
    \bm{\dot{x}_p}= f\left(\bm{x_p},u\right) = \bm{A_p}\bm{x_p}+\bm{B_p}\,{u}\\
	{y}= y_{p} = \bm{C_p}\, \bm{x_p}\\
\end{cases}
\end{IEEEeqnarray}
where $\bm{A_p = A + \bm{{\Phi}}}$. The vector $\bm{x_p}\in\mathbb{R}^{n_x\,\times\,1}$ represents $n_x$ states of the system, $u\in\mathbb{R}^{1}$ is the bounded and single control input to the system, $y\in\mathbb{R}^{1}$ is the bounded and single output state of the system, $\bm{A_p}\in\mathbb{R}^{n_x\,\times\,n_x}$ is the unknown and uncertain stabilizable system state matrix, $\bm{B_p}\in\mathbb{R}^{n_x\,\times\,1}$ is its unknown control input matrix, $\bm{C_p}\in\mathbb{R}^{1\,\times\,n_x}$ is its unknown output matrix and $\bm{\Phi}\in\mathbb{R}^{n_x\,\times\,1}$ is the unknown uncertainty (noise, and disturbance) vector affecting the system matrix $\bm{A}$.

Assume that this system can be represented as an input-output transfer function with unknown structure denoted as $P\left(s\right)$, where
\begin{IEEEeqnarray}{c}
P\left(s\right) = \frac{N_p\left(s\right)}{D_p\left(s\right)} = \frac{\sum_{i=0}^n{b_i}s^{n-i}}{\sum_{i=0}^n{a_i}s^{n-i}}
\end{IEEEeqnarray}
\figurename~{\ref{fig_prob}} shows a high-level diagram of this system connected in closed-loop to a PID controller (in software).

\subsection{Critic PID Control Form}\label{Scritic}
The control aim is to force the moving measured output $y$ of a dynamical system $P(s)$ to reach a desired and reachable output reference $r$ (setpoint). To do this, we need a computational algorithm that functions as a predictor of the control input $u$ to $P(s)$, that will asymptotically move $y$ close to $r$ in magnitude. Let us consider a simple critic computational form  for the PID, which is defined in (\ref{pideq1d}). (Note, the bracketed complex laplace operator notation $s$ will sometimes be removed from transfer functions and signals for simplicity.)
\begin{IEEEeqnarray}{c}
u=f(r,y)=\lambda_p\,u_p + \lambda_i\,u_i + \lambda_d\,u_d\IEEEyesnumber\label{pideq1d}\\
\noalign{\noindent where\vspace{\jot}}
u_p = K_p\,e_1 ,\quad u_i = K_i\,e_2 ,\quad u_d =K_d\,e_3 \\
e_1 = b\,r-y \equiv b\,r\left(s\right)-y\left(s\right) \\
e_2 = \int_0^{\tau}{\left( r-y \right)}\,dt = \int_0^{\tau}{e}\,dt \equiv \frac{e\left(s\right)}{s}\\
e_3 = {c\,}\dot{r}-\dot{y} \equiv s\left[c\,r\left(s\right)-y\left(s\right)\right]
\end{IEEEeqnarray}
The introduced terms $\lambda_p,\lambda_i,\lambda_d\,\in\,\mathbb{R^{+}}$ are termed the \enquote{critic} weights for each of the three PID contributing terms (or kernels) $u_p, u_i, u_d\in\mathbb{R}$ which are also related to the error terms $e_1, e_2, e_3\in\mathbb{R}$ respectively.

where $K_p\in\mathbb{R}$ is the proportional gain, $K_i\in\mathbb{R}$ is the integral gain, $K_d\in\mathbb{R}$ is the derivative gain, $e\in\mathbb{R}$ is the full error signal. This formulation is a 2DOF structure with respect to the proportional and derivative set-point limiters or weights $b$ and $c$ $\in\left[0,\,1\right]\in\mathbb{R}^{+}$

This critic formulation (\ref{pideq1d}) differs from the classical 1DOF or 2DOF PID definition. The classical PID form assumes that its three contributing terms equally dominate the output decision of the PID function, that is $\lambda_p,\lambda_i,\lambda_d\approx1$. As noted in \cite{hanPIDActiveDisturbance2009}, these are simplistic formulations of the natural error-based feedback control principle. Next, for the purpose of analysis, a compact system representation for the PID controller will be shown.

\subsection{Compact 2-DOF PID Representation}\label{Scomppid}
\begin{figure}[!t]
\centering
\includegraphics[width=2.5in]{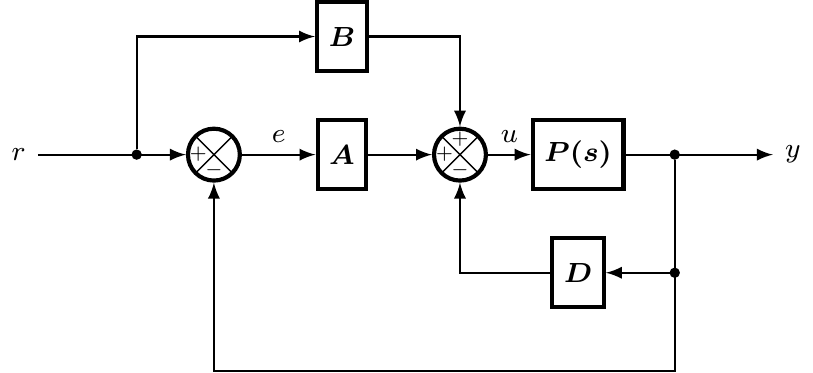}
\caption{Block diagram for the Closed-Loop PID Control System.}
\label{fig_pidtwodof}
\end{figure}
\begin{thm}
The two degree-of-freedom PID control law can be compactly represented as
\begin{IEEEeqnarray}{C}
u = B\;r + A\;e -D\;y\label{pidcomp}
\end{IEEEeqnarray}
where $B$ is the feedforward partition, $A$ is the integral error-feedback partition, and $D$ is the output-feedback partition of the PID control law.
\end{thm}
\begin{proof}
Without any loss of generality, assume that $\lambda_p,\lambda_i,\lambda_d=1$. Expanding (\ref{pideq1d}), then the expression (\ref{pideq2a}) is obtained. After further rearrangements, the result is the compact PID expression given by (\ref{pideq2c}). Observe that the signal $u$ has a distinct path to $r$ and $y$ as illustrated in the block diagram \figurename~{\ref{fig_pidtwodof}}.
\begin{IEEEeqnarray}{C}
\label{pideq2}
u = T_r\,r - T_y\,y \IEEEyesnumber\label{pideq2a}\\
\noalign{\noindent where\vspace{\jot}}
T_r = \left( K_p\,b+\frac{K_i}{s}+K_dsc\right)\\
T_y = \left( K_p+\frac{K_i}{s}+K_ds\right)\\
u=\left( K_p\,b+K_dsc \right) \;r+\frac{K_i}{s}\;e-\left( K_p+K_ds \right) \;y\label{pideq2b}\\
u = B\left(s\right)\,r\left(s\right) + A\left(s\right)\,e\left(s\right)-D\left(s\right)\,y\left(s\right)\label{pideq2c}\\
u = B\;r + A\;e -D\;y\IEEEnonumber\label{pideq2d}
\end{IEEEeqnarray}
The proof ends here.
\end{proof}
The PID control system design problem is then to use (\ref{pidcomp}) to appropriately regulate the moving output of the assumed unknown and uncertain dynamical system $P\left(s\right)$. This means forcing the dynamical system's output to closely follow and approach a desired and reachable output (set-point). In the next section, we will introduce the \enquote{Closed PID-Loop Model} Following Control (CPLMFC) design method and show that the PID's three main control parameters $K_p$, $K_i$, $K_d$ can be designed such that the output $y$ automatically follows and converges to the desired output $r$.

\section{CPLMFC Design Method}\label{sec4}
In this section, we show the main results for the CPLMFC, a method to automatically configure the main PID parameters. To start, first, we explicitly state some assumptions.

\subsection{Assumptions}\label{Sassump}
As regards the open-loop dynamic system $P\left(s\right)$, we make the following assumptions that are true for most known stabilizable dynamical systems in practise.
\begin{assum}\label{assump1}
Although all physical systems are nonlinear, $P(s)$ is designed to operate in a linear or an approximately linear input-output range.
\end{assum}
\begin{assum}\label{assump2}
$P(s)$ satisfies the bounded-input bounded-output (BIBO) definition of stability, which states that: a system is stable if every bounded input produces a bounded output.
\end{assum}
\begin{assum}\label{assump3}
If the dynamics of $P(s)$ is BIBO stable, according to Assumption~\ref{assump2}, then the unknown characteristic equation of $P(s)$ is Hurwitz (all its eigenvalues lie in the open left-half of the complex s-plane).
\end{assum}

\subsection{Closed PID-Loop Analysis}\label{Scpla}
To achieve the PID design goal, we start by obtaining the transfer function $T\left(s\right)$ of the overall closed loop system illustrated in \figurename~{\ref{fig_pidtwodof}}. Simplifying by block reduction, we have,
\begin{IEEEeqnarray}{c}
C=\frac{P\left(s\right)}{1+D\,P\left(s\right)}\label{redplanteq}
\end{IEEEeqnarray}
Again, by further block reduction operations using (\ref{redplanteq}), 
and then a signal-flow analysis using the Mason's Gain Formula (\ref{masoneq}),
where $G\left(s\right)$, $H\left(s\right)$, and $G\left(s\right)H\left(s\right)$ respectively represent the overall forward path gain, the overall feedback path gain, and the loop gain or transfer function of the PID loop in \figurename~{\ref{fig_pidtwodof}}. Finally, the closed PID-loop transfer function expression (\ref{pidplanttf}) is obtained.
\begin{equation}
T\left(s\right)=\frac{y\left(s\right)}{r\left(s\right)}=\frac{G\left(s\right)}{1+G\left(s\right)H\left(s\right)}=\frac{\sum_{i=1}^N{P_i\Delta _i}}{\varDelta}\label{masoneq}
\end{equation}
In this case, the number of forward paths from the input to output $N=2$. The $i$-th forward path gains $P_i$  are: $P_1=AC$ and $P_2=BC$, with a loop gain, $L_1=-AC$. Therefore, the signal flow determinant $\Delta=1-L_1=1+AC$; its co-factors (non-touching loops) along the $i_{th}$ forward path are: $\Delta_1=1$ and $\Delta_2=1$.
\begin{equation}
T = \frac{C\left(A+B\right)}{1 + {A}{}{C}}=\frac{\left(A+B\right)P}{1+\left(D+A\right)P}\label{pidplanttf}
\end{equation}
The analysis of the generic closed PID-loop transfer function (\ref{pidplanttf}) progresses to a proposition of  Theorem~\ref{loopgainthm}, the unity dominating loop-gain magnitude condition theorem.
\begin{thm}\label{loopgainthm}
The closed PID-loop system (\ref{pidplanttf}) can be approximated by the nominal \enquote{closed PID-loop model} (CPLM) (\ref{pidnm}), provided the unity dominating loop-gain magnitude condition (\ref{loopgaincond}) is satisfied.
\begin{IEEEeqnarray}{C}
\mid\frac{\left(A+D\right)\,N_p}{D_p}\mid\,\gg\,1\label{loopgaincond}\\
{T_m} = \frac{A+B}{D+A}\IEEEyesnumber\label{pidnm}
\end{IEEEeqnarray}
\end{thm}
\begin{proof}
To prove Theorem~\ref{loopgainthm}, we start by referring to the assumptions given in section~\ref{Sassump} which restricts the closed PID-loop characteristic equation $1+\left(D+A\right)P\left(s\right)$ given by the denominator of (\ref{pidplanttf}). Also, recall that $P = \frac{N_p}{D_p}$. Therefore, given that the Assumptions~\ref{assump1}--\ref{assump3} are satisfied, as $s\to0$, if the loop gain magnitude $\mid\frac{\left(A+D\right)\,N_p}{D_p}\mid$ of the closed PID-loop system becomes far greater than the value of one, then it is straightforward to see that
\begin{IEEEeqnarray}{c}
\lim_{s\rightarrow 0} {T} = \frac{\left( A+B \right) N_p}{D_p+\left( A+D \right) N_p} \approx \frac{A+B}{D+A}
\end{IEEEeqnarray}
This completes the proof.
\end{proof}
\begin{rem}\label{rem1}
Note that, when the magnitude of the loop gain is very much larger than unity under all conditions of interest. Then, the closed PID-loop output response is dominated by the PID parameters, hence can be regulated. This is similar in some sense, with H.S. Black's idea \cite{mbihi_analog_2018,thompson_intuitive_2013,astromAdvancedPIDControl2006} of the negative feedback amplifier system.
\end{rem}
For the closed PID-loop system (\ref{pidplanttf}), the approximated transfer function (\ref{pidnm}) is interpreted as the nominal \enquote{closed PID-loop model}. Consequently, this suggests an inherent implicit model following by the closed PID-loop system.

Also, a further analysis of (\ref{pidnm}), the nominal \enquote{closed PID-loop model} approximation of (\ref{pidplanttf}), shows that the control problem indirectly reduces the closed PID-loop system to follow the response of a second-order dominant closed-loop system. Accordingly, the design of (\ref{pidnm}) can be viewed as a dominant eigenvalue assignment problem. This can be posed as Theorem~\ref{domeig}.
\begin{thm}\label{domeig}
For the PID (with one derivative) control law, the closed PID-loop stabilizing objective implicitly defined by the \enquote{closed PID-loop model} can be approximately reduced to the dominant eigenvalue assignment of two complex poles on the left-hand side of the complex $s$-plane.
\end{thm}
\begin{proof}
From (\ref{pidnm}), $T(s)$ can be approximated to (\ref{pidmodel}). Also, the form of (\ref{pidmodel}) is equivalent to (\ref{pidmodelb}).
\begin{IEEEeqnarray}{C}\label{intpidplanttf}
\lim_{s\rightarrow 0} {T}\to{T_m} = \frac{A+B}{D+A}=\frac{cs^2+b\frac{K_p}{K_d}s+\frac{K_i}{K_d}}{s^2+\frac{K_p}{K_d}s+\frac{K_i}{K_d}}\label{pidmodel}\\
\frac{y}{r}\approx\frac{y_m}{r}\equiv\frac{cs^2+b2\zeta \omega _ns+\omega _{n}^{2}}{s^2+2\zeta \omega _ns+\omega _{n}^{2}}\label{pidmodelb}
\end{IEEEeqnarray}
where $\omega_n$ is the natural frequency, $\zeta$ is the damping factor and $y_m$ is the desired model output response designed to be followed with respect to $r$, by the actual closed PID-loop system output $y$. For an explicit model-following, the model output can be synthesized, without any loss of generality by setting both $b$ and $c$ to zero in the realization of (\ref{pidmodel}).

Since at steady state, we assume $T(s)$ is approximated by $T_m(s)$. If we fix the damping design parameter $\zeta=\frac{1}{\sqrt{2}}$ for a near optimal and robust output response of the uncertain dynamical system $P(s)$ \cite{yi_correcting_1989, i._j._nagrath_control_2006}. Then the dominant actual closed PID-loop eigenvalues of ${D_p+\left( A+D \right) N_p}$ will approximately converge to the nominal CPLM eigenvalues (\ref{xticeq}).
\begin{IEEEeqnarray}{c}
  s_{{1,}2}=-{\zeta}{\omega_n}\pm{j{\omega _n\sqrt{1-\zeta ^2}}}\\
  \mbox{Then: }s_{{1,}2} \approx -\frac{\sqrt{2}}{2}{\omega_n}\left(1\pm j\right)\label{xticeq}
\end{IEEEeqnarray}
This completes the proof.
\end{proof}
In the next section, we will use the CPLM design parameters, namely $\zeta,\omega_n$ to set a stabilizing $K_d$ and $K_i$ value of the PID controller.

\subsection{Setting the Integral and Derivative Gains}\label{Skikd}
The main result for setting the integral and derivative gains is posed as Theorem~\ref{titdset}
\begin{thm}\label{titdset}
The dominant eigenvalues of a closed PID-loop system can be approximated to the dominant eigenvalues of the CPLM, by setting both the integral and derivative time constants as (\ref{kdtd}) and (\ref{kiti}) respectively.
\begin{IEEEeqnarray}{c}
T_i=\frac{2\zeta}{\omega _n}=\frac{\sqrt{2}}{\omega_n}\IEEEyesnumber*\label{kiti}\\
T_d=\frac{1}{2\zeta \omega _n}=\frac{1}{{\sqrt{2}}\omega_n}\label{kdtd}
\end{IEEEeqnarray}
\end{thm}
\begin{proof}
We start by assuming Theorem~\ref{loopgainthm} is satisfied. This implies Theorem~\ref{domeig} is true. Comparing (\ref{pidmodel}) with (\ref{pidmodelb}), it can easily be deduced that the results are (\ref{kd}) and (\ref{ki}) for the respective derivative and integral gain parameters of the PID.
\begin{IEEEeqnarray}{c}
K_d=f_d\left( K_p,\omega _n,\zeta \right)=K_p\,T_d \IEEEyesnumber*\label{kd}\\
K_i=f_i\left( K_p,\omega _n,\zeta \right)=K_p\frac{1}{T_i}\label{ki}
\end{IEEEeqnarray}
The PID parameters (\ref{kdtd}) and (\ref{kiti}) are set using the natural frequency and damping factor, which is connected to the eigenvalue assignment and therefore stability of the nominal CPLM. Also the settings can be viewed as implicitly minimizing a quadratic cost function \cite{peterParameterAdaptiveControlBased1990}.
This completes the proof.
\end{proof}

In the next two sub-sections, the question of how to design the $\omega_n$ parameter of the CPLM will be addressed.

\subsection{CPLM: Normalized Response}\label{Snormresp}
\begin{figure}
\centering
\subfloat[]{\includegraphics[width=1in]{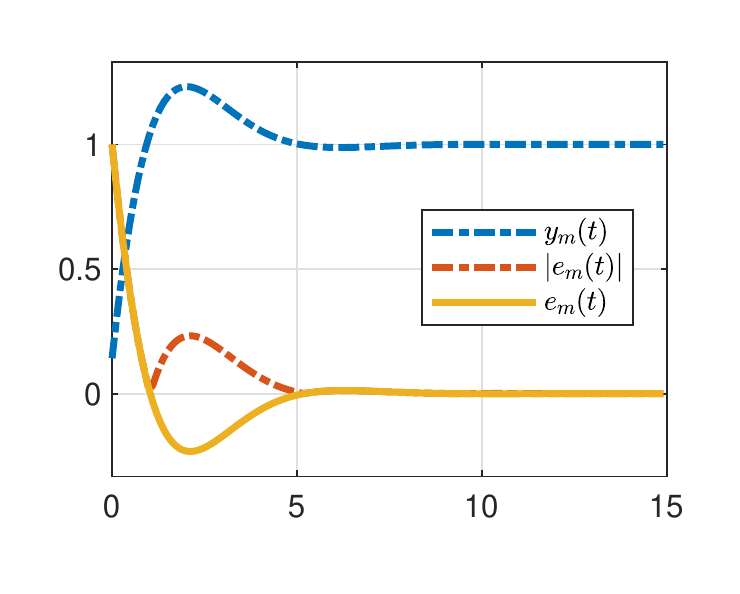}%
\label{fig0a}}
\quad
\subfloat[]{\includegraphics[width=1in]{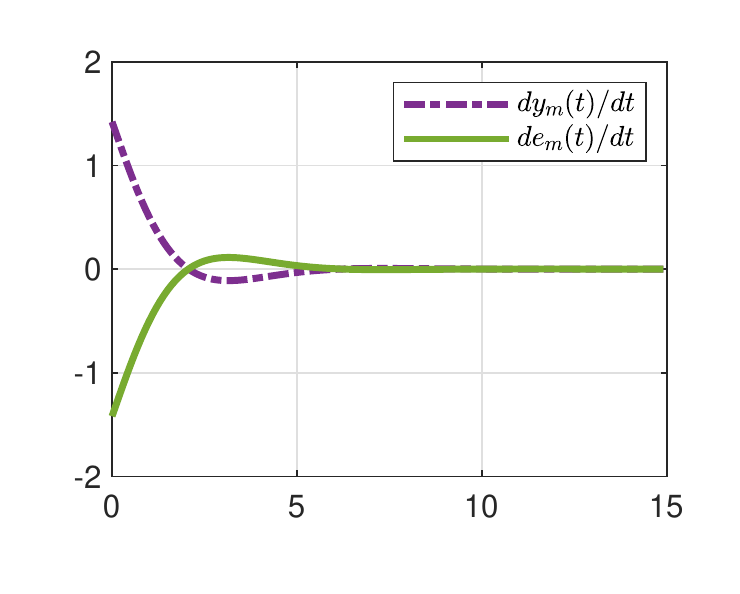}%
\label{fig0b}}
\quad
\subfloat[]{\includegraphics[width=1in]{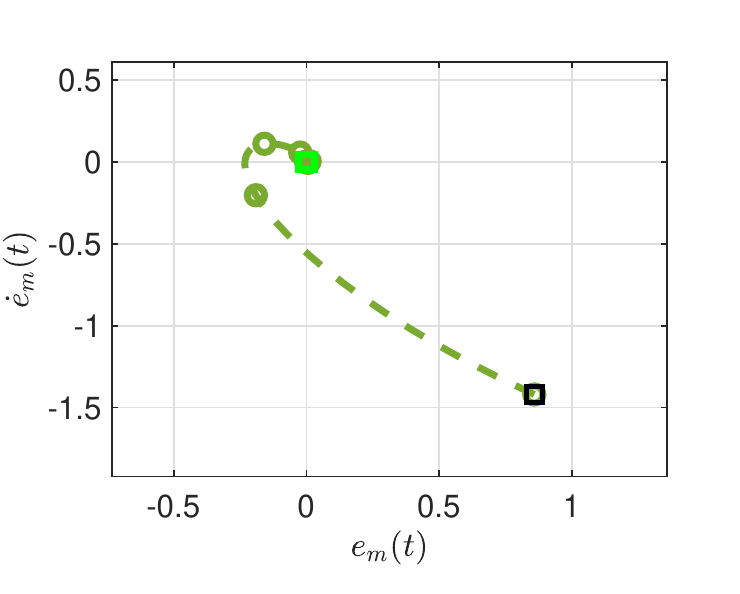}%
\label{fig0g}}
\quad
\subfloat[]{\includegraphics[width=1in]{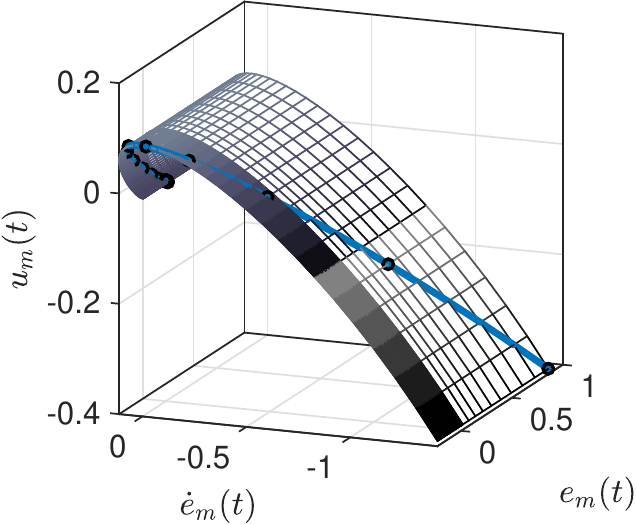}%
\label{fig0h}}
\caption{Closed PID-loop model's state space response analysis, when $b=1$, $c=0$; (\ref{fig0a}) output $y_m$ and error $e_m$, (\ref{fig0b}) output rate $\dot{y}_m$ and error rate $\dot{e}_m$, (\ref{fig0g}) error phase plot, and (\ref{fig0h}) control output surface map.}
\label{figpidm0}
\end{figure}

\begin{figure}
\centering
\subfloat[]{\includegraphics[width=1in]{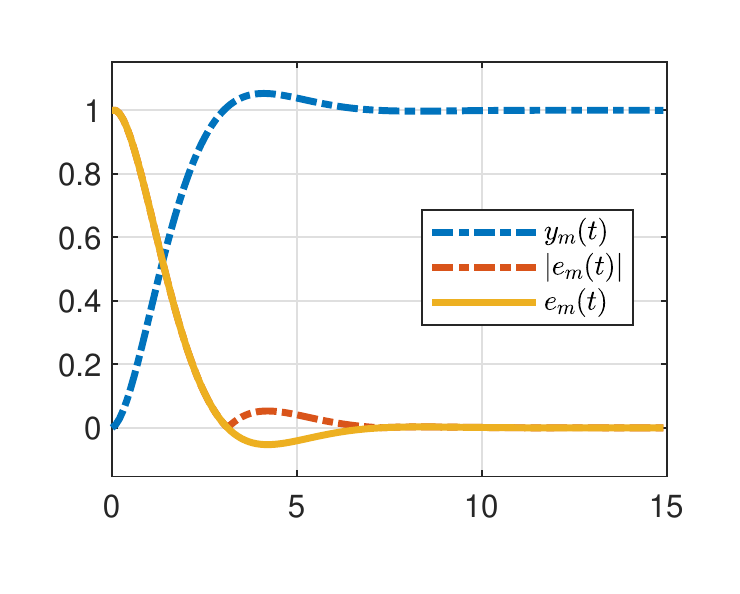}%
\label{fig1a}}
\quad
\subfloat[]{\includegraphics[width=1in]{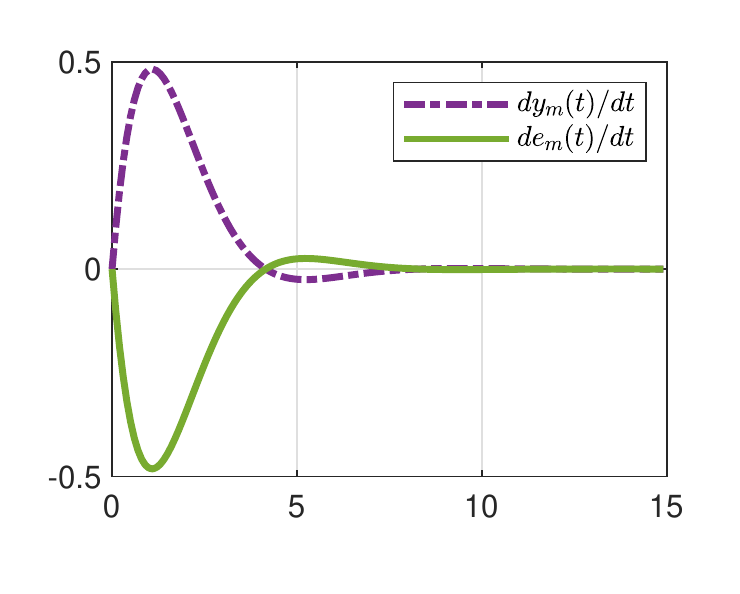}%
\label{fig1b}}
\quad
\subfloat[]{\includegraphics[width=1in]{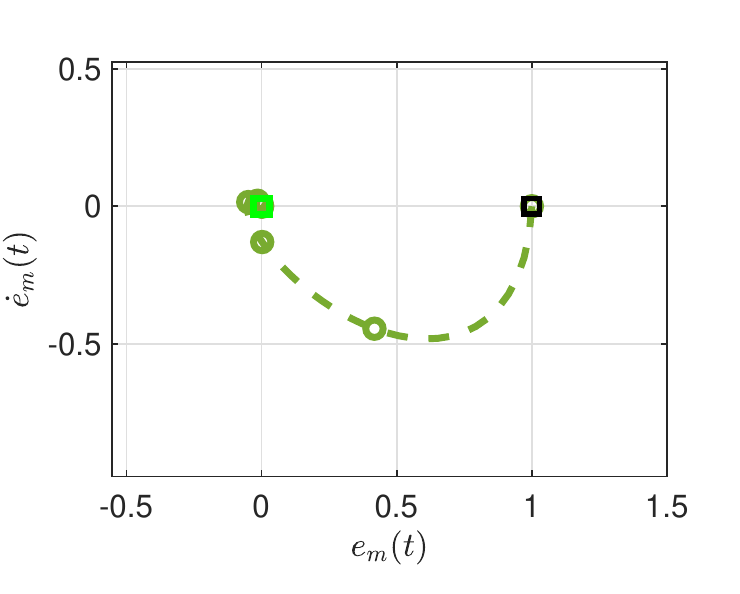}%
\label{fig1g}}
\quad
\subfloat[]{\includegraphics[width=1in]{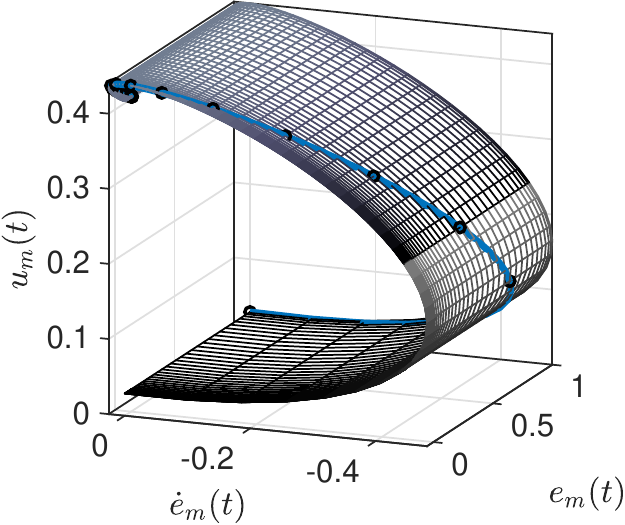}%
\label{fig1h}}
\caption{Closed PID-loop model's state space response analysis, when $b=0$, $c=0$; (\ref{fig1a}) output $y_m$ and error $e_m$, (\ref{fig1b}) output rate $\dot{y}_m$ and error rate $\dot{e}_m$, (\ref{fig1g}) error phase plot, and (\ref{fig1h}) control output surface map.}
\label{figpidm1}
\end{figure}

\begin{figure}[!t]
\centering
\subfloat[]{\includegraphics[width=1.5in]{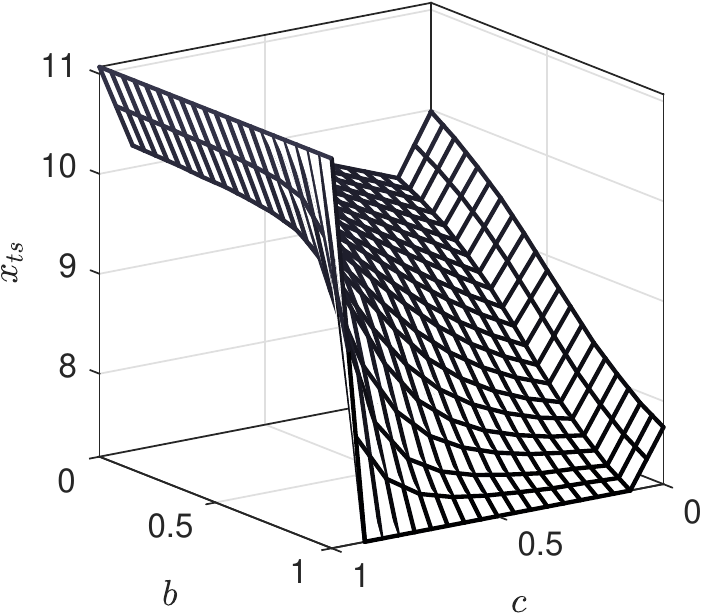}%
\label{figxts}}
\quad
\subfloat[]{\includegraphics[width=1.5in]{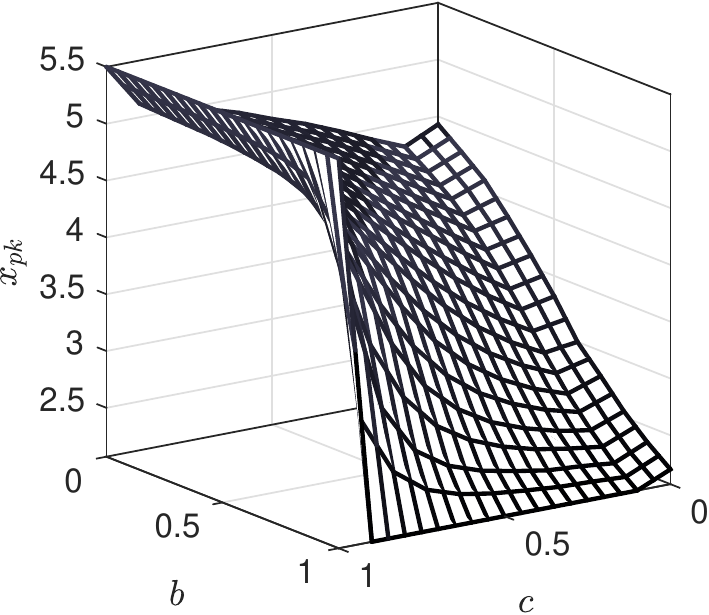}%
\label{figxpk}}
\caption{FIS output surface defined in Appendix~{\ref{fisapp}} for the normalized model settling and peak times.}
\label{figfuziosurf}
\end{figure}

In the previous section, the expressions (\ref{kiti}) and (\ref{kdtd}) depends on $\omega_n$. To design $\omega_n$, we start by analyzing the finite settling behaviour of the normalized nominal CPLM response ($\omega_n=1$) given by (\ref{pidmodelb}) for a constant or slowly varying unit-step input $r(s)=\textstyle{\frac{1}{s}}$. In observable canonical state-space form (\ref{ssmod}), the CPLM response $T_m\left(s\right)$ can be re-expressed as (\ref{sspidmodocf}).
\begin{IEEEeqnarray}{C}
\begin{array}{c}
\bm{\dot{x}_{m}}=\bm{A_m\,x_m}+\bm{B_m}r\\
{y_m}=\bm{C_m\,x_m}+\bm{D_m}r\\
\bm{x_m}=\left[ \begin{array}{c}
	x_{m1}\\
	x_{m2}\\
\end{array} \right] ,\bm{\dot{x}_m}=\left[ \begin{array}{c}
	\dot{x}_{m1}\\
	\dot{x}_{m2}\\
\end{array} \right]
\end{array}\label{ssmod}
\end{IEEEeqnarray}
where,
\begin{IEEEeqnarray}{c}\label{sspidmodocf}
\bm{A_m}=\left[ \begin{matrix}
	0&		-\omega _{n}^{2}\\
	1&		-2\zeta \omega _n\\
\end{matrix} \right]\,
\bm{B_m}=\left[ \begin{array}{c}
	\omega _{n}^{2}\left( 1-c \right)\\
	2\zeta \omega _n\left( b-c \right)\\
\end{array} \right]\\\IEEEnonumber
\bm{C_m}=\left[ \begin{matrix}
	0& 1\\
\end{matrix} \right]
\bm{D_m}=\left[ c \right]
\end{IEEEeqnarray}
\begin{IEEEeqnarray}{c}
y_m\left( t \right) = 1+e^{-\zeta \omega _nt} \left[
\begin{array}{l}
\left( c-1 \right) \cos \left( \omega _dt \right)+ \\
\left( 2b-c-1 \right) \zeta \frac{\omega _n}{\omega _d}\sin \left( \omega _dt \right)
\end{array}
\right]\label{ncplmresptime}
\end{IEEEeqnarray}
It is obvious that the CPLM response is affected by the choice of parameters $b$ and $c$. Changing the values of $b$ and $c$ will lead to variations in the normalized peak times and settling times of the CPLM. From, the analytic output response (\ref{ncplmresptime}), let the normalized time be \begin{IEEEeqnarray}{c}\label{xnom}
x={\zeta}{\omega_n}t
 \end{IEEEeqnarray}
Assume the normalized CPLM settling-time occurs at $x=x_s$ and its corresponding peak-time at $x=x_{pk}$. The simulated model response $T_m(s)$ is illustrated for two common forms: industrial form ($b=0, c=0$) in \figurename~{\ref{figpidm0}} and a general form ($b=1, c=0$) in \figurename~{\ref{figpidm1}} respectively.
The average of the first undershoot area of the normalized CPLM output response $y_m$ is considered to obtain $x_s$. For $x_{pk}$, the first point at which the time-derivative $\dot{y}_m=0$ is considered. The resulting normalized values for $x_{pk}$ and $x_{s}$ for standard finite values in the set $b$ and $c$ are shown in Table~\ref{xfis}.
\begin{table}[!t]
\centering
\begin{threeparttable}
\renewcommand{\arraystretch}{1.3}
\caption{Normalized Peak-Time\tnote{1} (Left) and Settling-Time\tnote{2} (Right): Closed PID-Loop Model's Fuzzy Membership Function Centers, $\bar{c}$}
\label{xfis}
\begin{tabular}{*{3}{c}||*{2}{c}}
\hline
\multirow{2}{*}{$c$} & \multicolumn{4}{c}{$b$}\\ \cline{2-5}
&0&1&0&1\\ \hline
0&4.43&2.20&9.98&7.74\\ \cline{1-5}
1&5.5&5.5&11.07&11.07\\ \hline
\end{tabular}
\begin{tablenotes}
\item[1] $\bar{d}=0.01$
\item[2] $\bar{d}=2.22$
\end{tablenotes}
\end{threeparttable}
\end{table}
The FSB of the nominal CPLM is quantified by its normalized settling-time $x_s$. With the aid of Table~\ref{xfis}, two approaches can be used in algorithm form to automatically compute the appropriate $x_{s}$ and  $x_{pk}$ with respect to the scalar weights $b$ and $c$ respectively. One, is to use a finite state machine (FSM) such as if-then rules or switch-case rules to encode the varying normalized values $x_s$ or $x_{pk}$. This approach is of low complexity and will work well for constrained processors. When processing requirements is not a concern, another approach is to use a fuzzy inference system (FIS) of fuzzy basis functions (FBF) \cite{mendelUncertainRuleBasedFuzzy2017} as is defined in Appendix~\ref{fisapp} to encode the knowledge of varying values of $x_s$ and $x_{pk}$ with respect to changing $b$ and $c$. This FIS design choice is outlined in Table~\ref{fischoice}, and the input-output nonlinear mapping surface of the FIS is shown in \figurename~{\ref{figfuziosurf}}.

Next, we will show with respect to expression (\ref{xnom}), how the normalized CPLM settling-time can be used to design the CPLM's $\omega_n$.

\subsection{CPLM: Natural Frequency Setting}\label{Swn}
It is straightforward to see from the normalized-time relationship (\ref{xnom}) that the CPLM's natural frequency can be designed by a rearrangement of variables. Changing $x=x_s$, we have:
\begin{IEEEeqnarray}{c}\label{wnid1}
\omega _n=f_{\omega_n}\left(\zeta,x_s,t_s\right)=\frac{x_{s}}{\zeta t_s}\label{wnid2}
\end{IEEEeqnarray}
where the CPLM natural frequency $\omega_n$ is matched to the dynamical system $P(s)$ by fixing $t$ as an identified closed-loop settling time $t=t_s$ of the physical system $P(s)$. This agrees with Assumption~\ref{assump2}. Also, since for every bounded input applied to $P(s)$, the resulting output does not instantly settle to a steady value (due to the concept of inherent input energy dissipation \cite{tewariModernControlDesign2002}). Therefore, through a quantifiable choice of the dynamical system's identified settling-time when in closed-loop, there is a natural FSB connection between the CPLM and the dynamical system. Next, we show how to identify this closed-loop settling-time.

\subsection{CPLM: Settling-Time Identification}\label{Stsident}
In the previous sub-sections, it was shown that in order to achieve closed-loop stability, the design of $\omega_n$ is system specific. The connection to settling-time was used to design the natural frequency of the CPLM. In this section, a procedure to identify this closed-loop settling-time of the system $P(s)$ is discussed.

\begin{figure}[!t]
\centering
\includegraphics[width=3.3in]{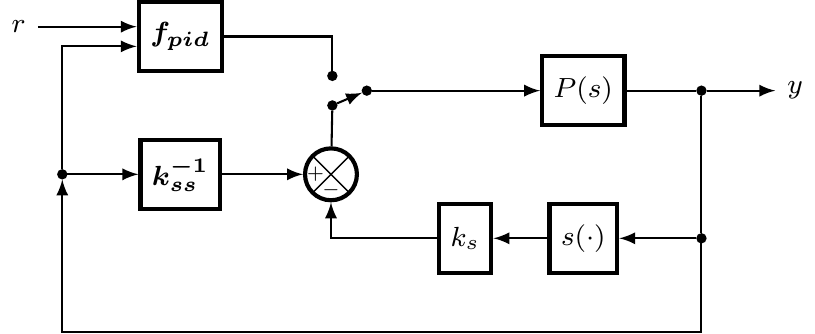}%
\caption{Finite Settling-Time Identification Process.}
\label{idents}
\end{figure}

We want to quantify the FSB of $P(s)$ that will be controlled in a closed PID-loop. To do this, identification of the closed-loop transient response is one appealing approach. It is known that a lot of engineering has gone into building physical systems that operate in an approximately linear range with a maximum input $u_{max}$ and output $y_{max}$. Therefore, the uncontrolled operation of such a system will be BIBO stable. To probe its FSB, this motivates the use of the inverse steady-state constant $k_{ss}^{-1}$ of the system to safely excite the system. An illustration for this identification is shown in Figure~\ref{idents}.
\begin{IEEEeqnarray}{c}\label{identseqn}
k_{ss}^{-1}=\frac{u_{max}}{y_{max}}\,;\quad u = k_{ss}^{-1}e - k_{s}\dot{y}
\end{IEEEeqnarray}
The closed-loop settling-time identification process is described by the following steps:
\begin{enumerate}[\IEEEsetlabelwidth{9}]
  \item  Select an appropriate sampling time $\tau$. 
  \item  Connect the closed-loop as defined by (\ref{identseqn}) and shown in Figure~\ref{idents} with a output-derivative state estimator.
  \item  Set the state estimator's logic as $k_s=0$
  \item  At initial time sequence, $n=0$, define the measured output as $y(0)= k\,y_{max}$
  \item  Set the input error as $e(n) = k\,y_{max} - y(n)$, where $0.5\le k \le1$.
  \item  Start the test and observe (and log) the output response $y(n)$.
  \item  Stop the identification test when an average steady state is reached (that is, when the difference in successive output samples consistently remains around a fixed point).
  \item  If oscillatory, set $k_s=1$. Restart from Step~6.
  \item  Else, use the logged output data to estimate the n-th sample $N_{\tau_l}$ at which the system starts to respond from rest, and the $n$-th sample $N_{ts}$ at which the system settles to an average stable value.
\end{enumerate}
The FSB of $P(s)$ can be quantified using $N_{ts}$, which is the system's nominal settling horizon in closed-loop. The identified settling-time can now be computed using,
\begin{IEEEeqnarray}{c}\label{tsN}
  t_s = T\left(N_{ts}-N_{\tau_l}\right)\\
  \tau_l = T\,N_{\tau_l} + \tau_c + \tau_y\label{Lcomp}
\end{IEEEeqnarray}
where $N_{ts}$ is the average number of sampled sequences it takes for the $y$ of a dynamical system $P(s)$ to settle to a steady value, while in a closed-loop. Also, $N_{\tau_l}$ is the delay horizon, the number of sampled sequences at which $y$ does not respond to the input $u$. Then the total delay-time, $\tau_l$ is computed as (\ref{Lcomp}). For an ideal sampled-data closed-loop system, without processing delays, $\tau_c$ and $\tau_y$ can be negligible. The delay-time $\tau_c$ can be obtained by timing the start to end of the PID computation. Also, the delay-time $\tau_y$ can be obtained by timing the end of the PID computation to the start of the next closed-loop sampled sequence.

The settling-time identification is important, as it determines the natural frequency of the CPLM, hence the closed PID-loop response stability of $P(s)$. To ensure the closed PID-loop response accuracy, the perfect model following error of the CPLM needs to be minimised for the closed PID-loop, while still maintaining stability. In the next section, a way to design the proportional gain $K_p$ to attempt to achieve this for the assumed unknown $P(s)$ is shown.

\subsection{CPLM: Adaptive Proportional Gain Settings}\label{Sadaptkp}

In this section, the goal is to determine a proportional gain $K_p$ in the unknown stabilizing and optimal set $\mathcal{K}$ of proportional gains $K_p^\star$. This is central to achieving a PMF of the CPLM. Note that, since $P(s)$ is assumed unknown and uncertain, this approach can only be an approximate method. It will be shown that a $K_p^\star \in \mathcal{K}$ can be asymptotically approached by an adaptive manner, and therefore minimize the PMF error.
Recall, the state-space equation for the system $P(s)$ is defined in (\ref{upltss1}). The desired model (implicit by default) is the CPLM. It's state-space equation is defined by (\ref{sspidmodocf}).
The main results are presented by the following Theorems.

\begin{thm}\label{pmfsufcond}
  The sufficient conditions for the closed PID-loop to achieve PMF of the CPLM are:
  \begin{enumerate}
    \item $\bm{B_m = B_p}\,K_p^\star\,e_c$
    \item $\bm{A_m = A_p - B_p\,C_p}\,K_p^\star\,e_c$
  \end{enumerate}
\end{thm}
\begin{proof}
First, we introduce some notations:
The critic-PID law can be rearranged as $u = u_p = {K_p}e_t$, where
\begin{IEEEeqnarray}{c}
e_t = e\left({\lambda_p + \lambda_i{T_i^{-1}}{s^{-1}} + \lambda_d{T_d}{s}}\right) = e\,e_c\\
e = r - y_p = r -\bm{C_p\,x_p}
\end{IEEEeqnarray}
The tracking error dynamics is $\bm{\dot{e}=\dot{x}_m -\dot{x}_p}$. The system state-equation can be rewritten as
\begin{IEEEeqnarray}{c}
\bm{\dot{x}_p} = \left(\bm{A_p-B_p\,C_p}\,K_p\,e_c\right)\,\bm{x_p} + \bm{B_p}\,K_p\,e_c\,r
\end{IEEEeqnarray}
Therefore the tracking error dynamics is
\begin{IEEEeqnarray}{Rl}\label{trerrdyn}
\bm{\dot{e}} = & \bm{A_m\,x_m} + \bm{B_m}\,r - \IEEEnonumber\\
&\left(\bm{A_p-B_p\,C_p}\,K_p\,e_c\right)\,\bm{x_p} - \bm{B_p}\,K_p\,e_c\,r%
\end{IEEEeqnarray}
At PMF, $K_p \to K_p^\star$, and therefore, $\bm{x_p \to x_m}$, and $\bm{\dot{e}} = 0$. Then (\ref{trerrdyn}) can be rearranged as
\begin{IEEEeqnarray}{Rl}
&\bm{A_m\,x_m} + \bm{B_m}\,r = \IEEEnonumber\\
&\left(\bm{A_p-B_p\,C_p}\,K_p^\star\,e_c\right)\,\bm{x_m} + \bm{B_p}\,K_p^\star\,e_c\,r%
\end{IEEEeqnarray}
Comparing both sides, we have:
\begin{IEEEeqnarray}{C}
\bm{B_m = B_p}\,K_p^\star\,e_c\\
\bm{A_m = A_p - B_p\,C_p}\,K_p^\star\,e_c
\end{IEEEeqnarray}
This concludes the proof.
\end{proof}

\begin{thm}\label{stabthm}
The closed PID-loop system will be asymptotically stable and accurately follow the CPLM, if and only if
the condition (\ref{stabcnd}) is satisfied.
\begin{IEEEeqnarray}{c}\label{stabcnd}
\lim_{t\rightarrow \infty} \bm{\bar{A}_{pc}} \to \bm{A_m}
\end{IEEEeqnarray}
where $\bm{\bar{A}_{pc}}$ represents the dominating eigenvalue matrix in the closed PID-loop eigenvalue matrix $\bm{A_{pc}} = \bm{A_p-B_p\,C_p}K_p\,e_c$.
\end{thm}
\begin{proof}
First, recall that the CPLM matrix $\bm{A_m}$ is hurwitz, so its eigenvalues strictly lie in the left-half of the complex $s$-plane. The closed PID-loop system can be represented as:
\begin{IEEEeqnarray}{c}\label{clxmxp}
\left[ \begin{array}{c}
	\bm{\dot{x}_m}\\
	\bm{\dot{x}_p}\\
\end{array} \right] =\left[ \begin{matrix}
	\bm{A_m}&		0\\
	0&		\bm{A_{pc}}\\
\end{matrix} \right] \left[ \begin{array}{c}
	\bm{x_m}\\
	\bm{x_p}\\
\end{array} \right] +\left[ \begin{array}{c}
	\bm{B_m}\\
	\bm{B_{pc}}\\
\end{array} \right] r
\end{IEEEeqnarray}
where $\bm{B_{pc}} = \bm{B_p}K_p\,e_c$. Hence, the closed PID-loop system (\ref{clxmxp}) will be asymptotically stable and accurately follow the CPLM (\ref{sspidmodocf}), if and only if $\bm{A_{pc}}$ the dominating eigenvalues in the closed PID-loop matrix approach the eigenvalues of $\bm{A_m}$.
This completes the proof.
\end{proof}

\begin{thm}\label{kpupdthm}
 The proportional gain $K_p$ can be chosen through an adaptive rule
\begin{IEEEeqnarray}{c}\label{kpupd}
\dot{K}_p= \alpha\,\gamma\,e{e_t}
\end{IEEEeqnarray}
 in order to minimise the $\mathcal{L}_2$ norm of the PMF error defined by
\begin{IEEEeqnarray}{c}\label{pmferrcnd}
{\lVert{\bm{\dot{e}}-\bm{A_m\,e}}\rVert}_2={\lVert{\bm{B_p}\left( K_{p}^{\star}-K_p \right) e_t}\rVert}_2
\end{IEEEeqnarray}
\end{thm}
\begin{proof}
Note that $\bm{x_m = e + x_p}$, so the tracking error dynamics (\ref{trerrdyn}) can be rearranged as:
\begin{IEEEeqnarray}{Rl}
\bm{\dot{e}} = & \bm{A_m\,e} + \left(\bm{B_m}-\bm{B_p}\,K_p\,e_c\right)r \IEEEnonumber\\
&+\:\left(\bm{A_m-A_p} + \bm{B_p\,C_p}K_p\,e_c\right)\bm{x_p}%
\end{IEEEeqnarray}
At PMF, Theorem~\ref{pmfsufcond} is satisfied, so (\ref{trerrdyn}) becomes
\begin{IEEEeqnarray}{Rl}
\bm{\dot{e}} = \bm{A_m\,e} + \bm{B_p}\,e_\mathcal{K}\,e_t\\
\mbox{where,}\quad e_\mathcal{K} = \left(K_p^\star -K_p\right)%
\end{IEEEeqnarray}
Now, a candidate positive definite lyapunov function $V$ of $e$ and $e_\mathcal{K}$ can be constructed such that
\begin{IEEEeqnarray}{C}
V\left(e,e_\mathcal{K}\right) = \frac{\alpha}{2}e^2 + \frac{\bm{B_p}}{2\,\gamma}e_{\mathcal{K}}^2
\end{IEEEeqnarray}
and its first derivative and second derivative with respect to time are:
\begin{IEEEeqnarray}{RRl}
\dot{V} &= \alpha\bm{A_m}e^2 +\bm{B_p\,}e_\mathcal{K}\left( \alpha\,e\,e_t - \frac{\dot{K}_p}{\gamma}\right)\\
& \ddot{V} = \dot{e}\left(2\,\alpha\bm{A_m}e + \bm{B_p}\,\alpha\,e_\mathcal{K}\,e\,e_t\right)\IEEEnonumber\\
&-\: \bm{B_p}\,\dot{K}_p\left(\alpha\,e\,e_t - \frac{\dot{K}_p}{\gamma}\right)%
\end{IEEEeqnarray}
Note that, if $\dot{K}_p= \alpha\,\gamma\,e{e_t}$ is selected to cancel out the second-term of $\dot{V}$, then $\dot{V}$ becomes negative semi-definite. By virtue of $\dot{V}\le0$, then $e_\mathcal{K}$,$e$ and $e_t$ will be bounded. Also, this choice makes $\ddot{V}$ bounded, meaning that $\dot{V}$ is uniformly continuous in time. Therefore, it follows from the application of Babarlat's Lemma \cite{tien2002sliding} that $\dot{V} \to 0$, $e \to 0$ as $t\to \infty$. This implies that the tracking error is asymptotically stable.

Since $K_p$ is only shown to be bounded according to $e_\mathcal{K}$, this choice does not guarantee the accuracy of the adaptive system. It follows that Theorem~6 is a sufficient condition for asymptotic stability and accuracy. Consequently, it is straightforward to see that if $K_p{\nrightarrow}K_p^\star$, then the adaptive update (\ref{kpupd}) will at least minimize the $\mathcal{L}_2$ norm of the PMF error, that is: ${\lVert{\bm{\dot{e}-A_m\,e}\rVert}_2=\lVert{\bm{B_p}\,e_{\mathcal{K}}\,e_t}\rVert}_2$%
.
This completes the proof.
\end{proof}%
Therefore, at every sampling instance, the adaptive update rule for $K_p$ can be re-expressed as:
\begin{IEEEeqnarray}{C}\label{kpadapt}
K_p = K_p + \dot{K}_p\\
\dot{K}_p = \alpha\,\gamma\,e\,\min\left(u_{max}, \left|e_t\right|\right)
\end{IEEEeqnarray}
where $\alpha\in\mathrm{R}^+$ and $\gamma\in\mathrm{R}$. $\alpha$ is a tunable constant that tries to achieve $e_\mathcal{K}\to0$, while $\gamma=0.001$ is a small weighting constant.
$\\$%

The adaptation given by (\ref{kpadapt}) does not consider the presence of delay-time present in the closed PID-loop. When the magnitude of the total delay-time gets larger, the control input $u$ becomes evidently delayed, and the solution obtained from Theorem~\ref{kpupdthm} above cannot alone guarantee that the closed PID-loop will follow the CPLM. Therefore, a faster way to initialize and reach $K_p^\star$ is needed. But the optimal stabilizing set $\mathcal{K}$ is unknown. To solve this problem, a fictitious upper limit for $K_p$ such that $0 < K_p < k_{plim}$ is constructed. We define $k_{plim}$ as:
\begin{IEEEeqnarray}{c}\label{kplim}
k_{plim}= \alpha\,(\kappa_g)\,\frac{\tau_l+t_s}{t_s}
\end{IEEEeqnarray}
In this test case, $k_{plim} \triangleq \kappa_g$. That is $\kappa_g$ becomes a free tunable parameter. The aim is to obtain a stabilizing and accurate $k_{plim}$ such that an initial guess value $K_{p0}$ can be obtained from a nonlinear function of the current error $e$ and current predicted control-input $u$. This function is defined by using $\mathrm{nlsig}^-$ the forward form of the n-logistic sigmoid function, with at least $n=1$, defined in Appendix~\ref{nlsigapp}.
\begin{IEEEeqnarray}{Cr}\label{kp0}
K_{p0} & = \mathrm{nlsig}^-(e,x(e),-x(e),k_{plim},-k_{plim},\IEEEnonumber\\
& 1,0.1,0) + \mathrm{nlsig}^-(u,x(u),-x(u),\IEEEnonumber\\
& k_{plim},-k_{plim},1,0.1,0)\\
& \mbox{where,}\quad x(\star) = k_{plim} + \star
\end{IEEEeqnarray}
This value is then further penalised to be within bounds using
\begin{IEEEeqnarray}{C}\label{Kp}
K_p = \mathrm{nlsig}^-(K_{p0},x(K_{p0}),0,k_{plim},0,n,0.1,0)
\end{IEEEeqnarray}
Then, an experimental closed PID-loop response of a normalized first-order dynamics with increasing normalized real-valued delay-time $\tau_l$ in the normalized set $\left[0, 0.01, 0.1, 0.5, \cdots, 20, 30\right)$ was simulated. Through curve-fitting, the experimental data was then used to correlate the values of $\tau_l$ and the resulting
$k_{plim}= \left[2, 2, 1.5, 0.8, \cdots, 0.12, 0.1 \right)$ that led to a stabilizing $K_p$ value within the constrained limiting interval. The final result is a rational function (\ref{ratfitKp}) expressed below.
\begin{IEEEeqnarray}{c}
\kappa _g=\frac{p_1\bar{\tau}_l^2+p_2\bar{\tau}_l+p_3}{\bar{\tau}_l^2+q_1\bar{\tau}_l+q_2}\label{ratfitKp}\\
\noalign{\noindent such that,} \bar{\tau_l}=\frac{\tau_l-5.936}{8.771}
\end{IEEEeqnarray}
where the coefficients are $p_1=0.05132$, $p_2=0.2041$, $p_3=0.1214$, $q_1=1.538$, $q_2=0.5864$. Finally, we have:
\begin{IEEEeqnarray}{C}\label{kpfinal}
K_p = f_p\left(e,u,e_t,\alpha,\tau_l,t_s\right)
\end{IEEEeqnarray}
The ordered arrangement inside the proportional gain function (\ref{kpfinal}) is such that: (\ref{ratfitKp}) (\ref{kplim}), (\ref{kp0}), (\ref{Kp}) are executed, and then (\ref{kpadapt}) is not executed until the current time is greater than the estimated delay-time from (\ref{Lcomp}).
\begin{figure}[!t]
\centering
\includegraphics[width=2.5in]{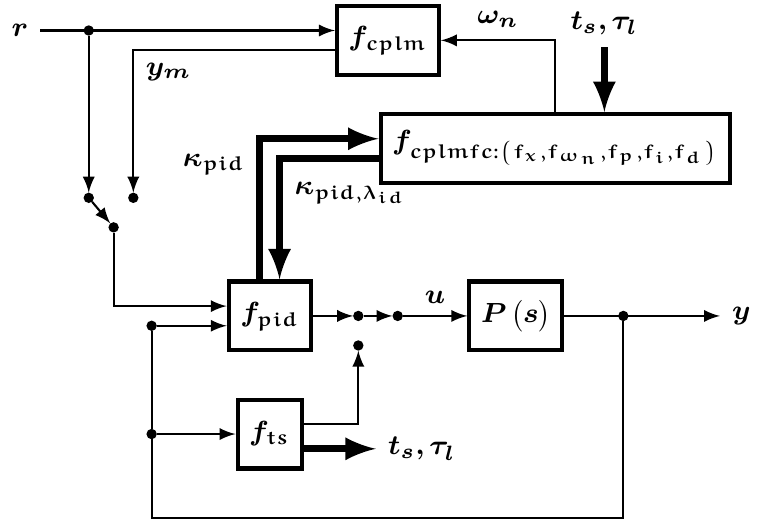}%
\caption{Overview of the Closed PID-Loop Model Following Control (CPLMFC) Method.}
\label{ovd}
\end{figure}

Now the three main PID parameters have been systematically configured. An overview of the CPLMFC method is illustrated in \figurename~\ref{ovd}. In the next section, an automatic tuning algorithm based on this method will be presented.

\subsection{CPLMFC: Algorithm Implementation}\label{Salgcplmfc}
The implementation of the CPLMFC algorithm inside the sampled closed PID-loop is illustrated in Algorithm~\ref{algcplmfc} with the main gains set using the formulas derived in the previous subsections of this section. On the other hand, since the aim of criticism is to fine-tune decisions, the critic weights can be viewed as fine-tuning knobs. They are set manually using some recommended conditions given in Algorithm~\ref{reccritic}.

The proportional critic weight $\lambda_p$ should be fixed to unity. It represents how much belief is placed on current experience of error. $\alpha$ controls the response speed, how fast or slow, $y \to y_m$. A default starting value is $\alpha=1$. $\lambda_i$ represents how much belief is placed on accumulated error experience, and controls the accuracy of $y \to y_m$. In normal cases of regulation or tracking, $0 \le \lambda_i \le 1$. The value can then be increased or decreased in this range depending on simulations. The lower the value the lower the overshoot. On the other hand, $\lambda_d$ represents how much belief is placed on future error estimations or projections. It can dampen or increase the oscillations in $y$. In normal cases of regulation or tracking, $0 \le \lambda_d \le 0.5$. For stable systems, the only value that may need to be increased or decreased is $\alpha$.

\begin{algorithm}[]
\caption{CPLMFC Tuning Algorithm}\label{algcplmfc}
\begin{algorithmic}[1]
\Require $\tau,\tau_l,t_s,e,u,b,c,\zeta$
\State $\left[x_s,x_{pk}\right] \gets f_{x}\left(b,\,c\right)$
\State $\omega_n \gets f_{\omega_n}\left(\zeta,x_s,t_s\right)$
\Ensure $\alpha > 0$  and at $t = 0: K_p = 0.01, K_i = K_d = 0$
\State $K_p \gets f_p\left( e\left(t\right),u\left(t\right),\alpha,\tau_l,t_s\right)$
\State $K_i \gets f_i\left( K_p,\omega _n,\zeta \right)$
\State $K_d \gets f_d\left( K_p,\omega _n,\zeta \right)$
\end{algorithmic}
\end{algorithm}
In the next section, the CPLMFC method will be applied first to the numerical simulation of a normalized third-order dynamical system and then a more challenging case in form of a second-order (uncertain, non-linear and integrating) dynamical system is considered as a representative case-study.

\begin{algorithm}[]
\caption{Critic Settings}\label{reccritic}
\begin{algorithmic}[1]
\State $\lambda_p \gets 1$
\Ensure $1 \ge \lambda_i^2\ge\lambda_i^1 > 0$;
\If{$\tau_l>\tau$}
\State $\lambda _i \gets \lambda_i^1$
\Else
\State $\lambda _i \gets \lambda_i^2$ 
\EndIf
\Ensure $1 \ge \lambda_d^2\ge\lambda_d^1 \ge 0$; 
\If{$\tau_l>\tau$}
\State $\lambda _d \gets \lambda _d^1$
\Else
\State $\lambda _d \gets \lambda_d^2$ 
\EndIf
\end{algorithmic}
\end{algorithm}

\section{Simulations}\label{sec5}
In the previous section, using the settling-time identification of a dynamical system, a CPLMFC algorithm was designed for real-time robust optimization of the response of a closed PID-loop system consisting of a critic 2DOF PID control law and the dynamical system assumed to be unknown. Further, in this section, some simulation results will be presented and discussed.

In this section, for standard measure of closed-loop performance, the integral absolute error index $J_{iae}$, and (or) the integral absolute error index $J_{iae}$ are used to measure the error regulation.
\begin{IEEEeqnarray}{c}
J_{iae}=\int_0^{\infty}{\left| e\left( t \right) \right|}\quad,\quad J_{ise}=\int_0^{\infty}{e^2\left( t \right)}
\end{IEEEeqnarray}
Also, other common and intuitive indicators of the quality of a closed-loop control system, is the performance in terms of the resulting maximum overshoot and settling time ($1\%$) \cite{bucz_advanced_2018,wang_approach_2018,jantzen_turning_2016-1,viteckova_2dof_2015-1}.

\subsection{Stable, Delay-Free System}
The normalized third-order plant transfer function model is widely considered as a benchmark plant for evaluating PID control design methods. It is a representative plant for some stable physical dynamical systems employed in the industry \cite{astromBenchmarkSystemsPID2000}.
\[
P\left( s \right) =\frac{1}{(s+1)^3}
\]
Selecting $b=1$, $c=0$, maximum control input $u_{max} = 10$, and $\tau =\textstyle{\frac{1}{10}}$secs. The closed PID-loop settling-time identification of $P(s)$ is first carried out, the settling horizon is selected as $N_{ts}=100$ and in the case $N_{\tau_l} = \tau_c = \tau_y = 0$.

\begin{figure}[!t]
\centering
\subfloat[]{\includegraphics[width=2in]{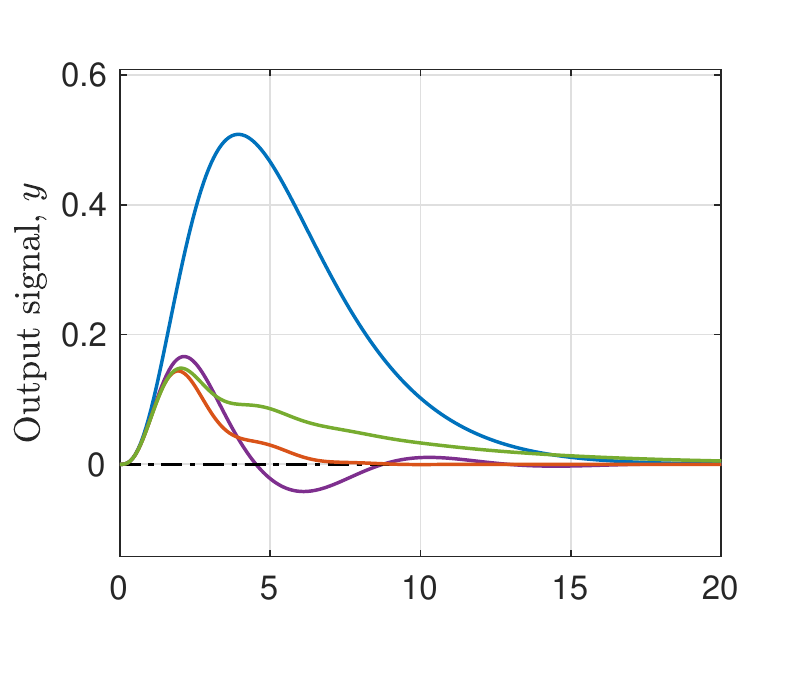}%
\label{figp3aa}}
\caption{Responses to a unit step load disturbance for $P\left(s\right)$: (convex-concave IAE optimization (solid purple), CPLMFC (($\alpha=2$, solid blue), ($\alpha=16$, solid red), ($\alpha=16,\lambda_i=0.25$, solid green) closed PID-loop model response (dash-dotted black)).}
\label{figp3a}
\end{figure}

\begin{figure}[!t]
\centering
\subfloat[]{\includegraphics[width=2in]{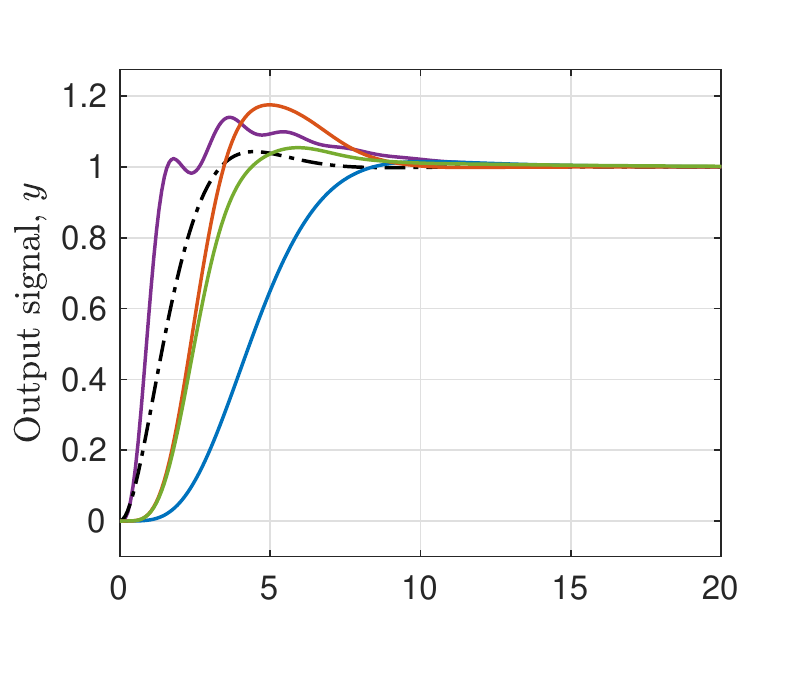}%
\label{figp3ba}}
\caption{Responses to a unit step set-point command, no disturbance for $P\left(s\right)$: (JJST (solid purple), CPLMFC (($\alpha=2$, solid blue), ($\alpha=16$, solid red), ($\alpha=16,\lambda_i=0.25$, solid green) closed PID-loop model response (dash-dotted black)).}
\label{figp3b}
\end{figure}

\begin{figure}[!t]
\centering
\subfloat[]{\includegraphics[width=2in]{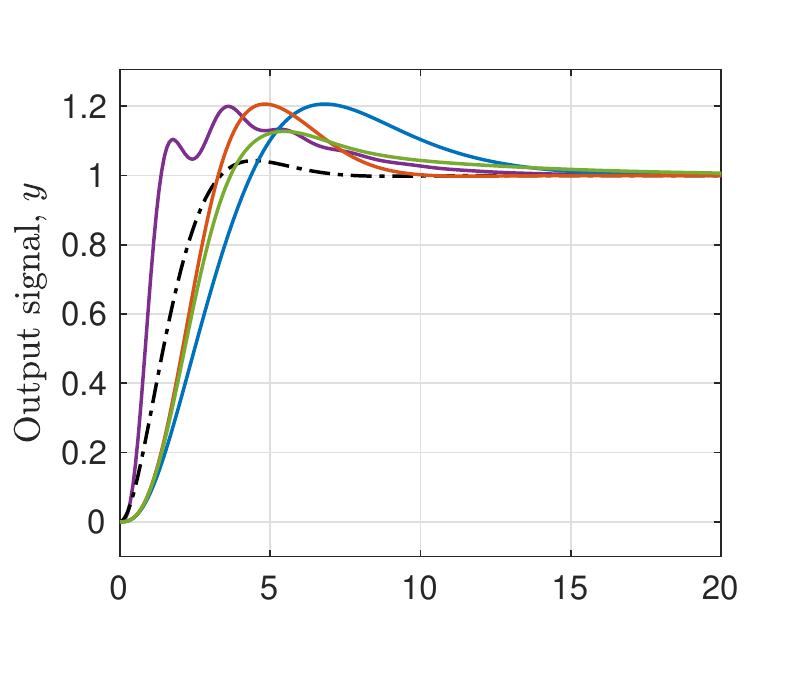}%
\label{figp3ca}}
\caption{Responses to a unit step set-point command, with unit-step input load disturbance for $P\left(s\right)$: (JJST (solid purple), CPLMFC (($\alpha=2$, solid blue), ($\alpha=16$, solid red), ($\alpha=16,\lambda_i=0.25$, solid green) closed PID-loop model response (dash-dotted black)).}
\label{figp3c}
\end{figure}

\begin{figure}[!t]
\centering
\subfloat[]{\includegraphics[width=1.65in]{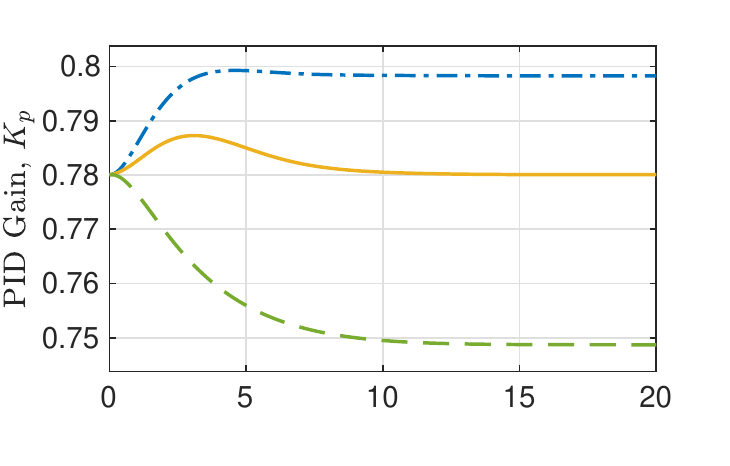}%
\label{figpKpa}}
\subfloat[]{\includegraphics[width=1.65in]{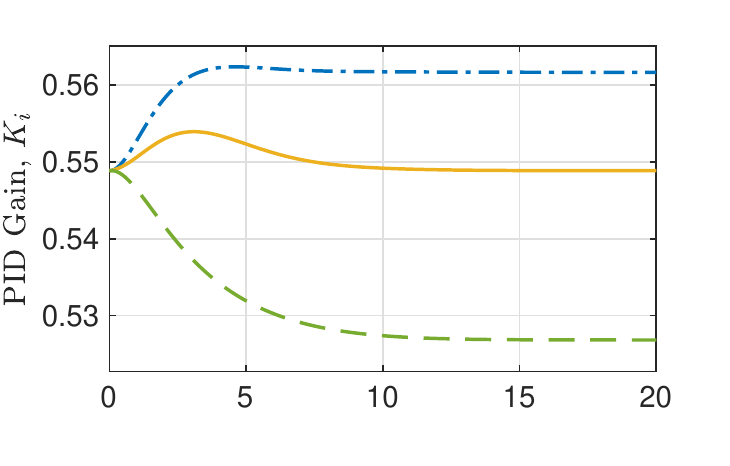}%
\label{figpKpb}}
\quad
\subfloat[]{\includegraphics[width=1.65in]{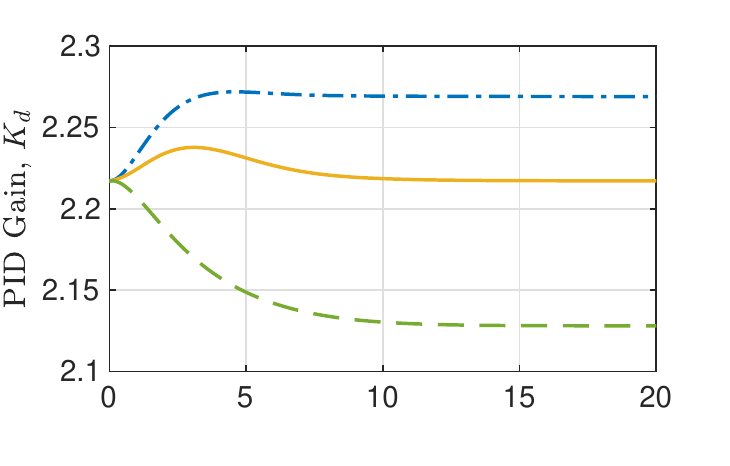}%
\label{figpKpc}}
\quad
\subfloat[]{\includegraphics[width=1.65in]{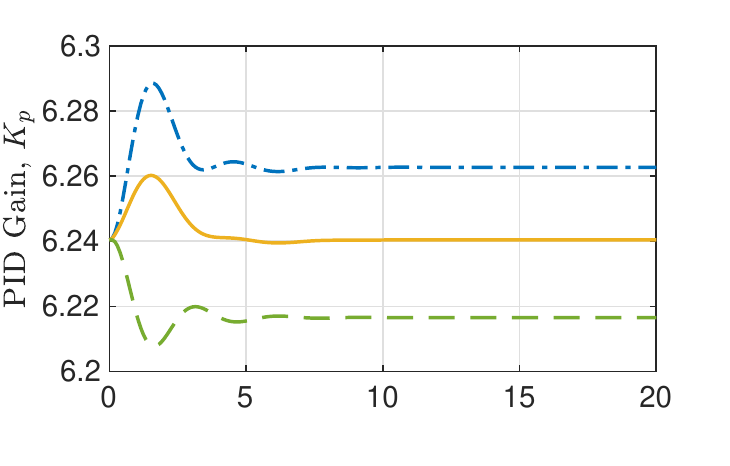}%
\label{figpKpd}}
\subfloat[]{\includegraphics[width=1.65in]{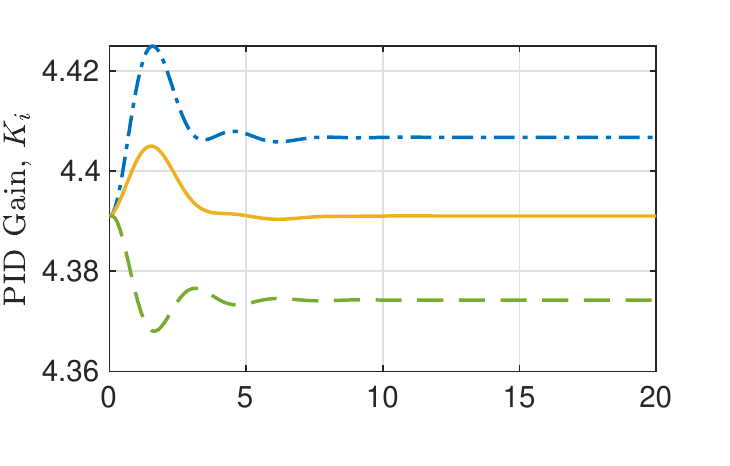}%
\label{figpKpe}}
\quad
\subfloat[]{\includegraphics[width=1.65in]{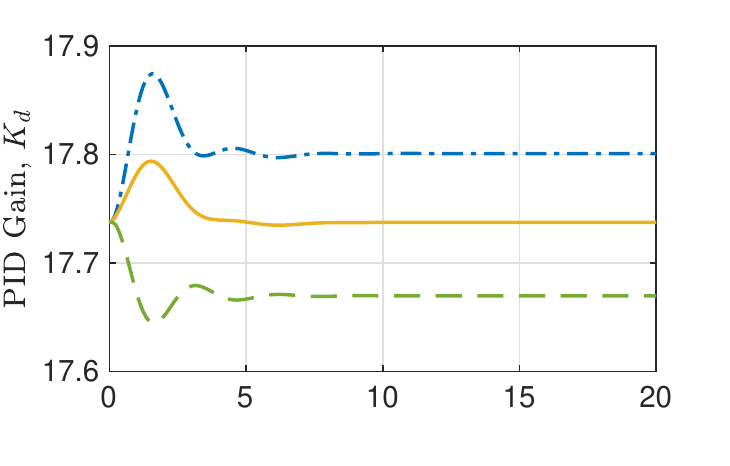}%
\label{figpKpf}}
\caption{CPLMFC Adaptive Gains Evolution: Top Three ($\alpha=2$) Bottom Three ($\alpha=16$); unit-step with no disturbance (dash-dotted blue), unit-step with unit-step load disturbance (solid brown), unit-step input load disturbance (dashed green). }
\label{figKp}
\end{figure}

The performance of the CPLMFC method is compared against two relatively recent methods in the literature, namely: the Convex-Concave (CC-IAE) optimization method \cite{hastPIDDesignConvexconcave2013} for load disturbance regulation performance and Jantzen-Jakobsen's Settling Time (JJST) method \cite{jantzen_turning_2016-1} for set-point regulation performance. Shown in \figurename~\ref{figp3a}, \figurename~\ref{figp3b} and \figurename~\ref{figp3c} are the controlled output responses to a unit step input load disturbance, a unit-step setpoint command, and a unit-step set-point command with unit-step disturbance respectively.
The adaptive evolution of the PID gains during the closed loop operation is shown in \figurename~\ref{figKp}.
It can be concluded from the performance measures in Table~\ref{pmeasureP2} that the CPLMFC method exhibits comparable performance with the other methods. On average, it results in a very good compromise between the transient properties of  maximum overshoot, final settling-time ($1\%$) and the $J_{iae}$ index.
\begin{table*}[!t]
\centering
\footnotesize{
\begin{threeparttable}
\renewcommand{\arraystretch}{1.3}
\caption{Controller Parameters and Performance Measures for the Plant $P=\frac{1}{(s+1)^3}$.}
\label{pmeasureP2}
\centering
\begin{tabular}{*{10}{c}}
\toprule
& $J_{iae}$& $t_s(1\%)$ & $y_{max}$& $K_p$& $K_i$& $K_d$& $\lambda_i$& $\lambda_d$ \\
\midrule
CC-IAE\tnote{1}& 0.529& 10.77& 0.17& 3.81& 3.33& 4.25& 1&1 \\ \hline
CPLMFC ($\alpha=2$)\tnote{1}& 3.127& 15.16& 0.51& 0.75& 0.53& 2.13& 0.6& 0.25\\ \hline
CPLMFC ($\alpha=16$)& 0.383& 6.19& 0.14& 6.22& 4.37& 17.67& 0.6& 0.25\\ \hline
CPLMFC ($\alpha=16$)& 0.886& 16.48& 0.15& 6.22& 4.37& 17.67& 0.25& 0.25\\ \hline
\midrule
JJST (R=1)\tnote{2} & 1.519& 11.94& 1.14& 10& 2.70& 9.26& 1&1 \\ \hline
CPLMFC ($\alpha=2$)\tnote{2}& 3.12& 12.39& 1.02& 0.8& 0.56& 2.27&0.6& 0.25 \\ \hline
CPLMFC ($\alpha=16$)\tnote{2}& 1.344& 9.26& 1.18& 6.26& 4.41& 17.8&0.6& 0.25 \\ \hline
CPLMFC ($\alpha=16$)\tnote{2}& 1.269& 10.35& 1.05& 6.26& 4.41& 17.8&0.25& 0.25 \\ \hline
\midrule
JJST (R=1)\tnote{3} & 1.773& 12.49& 1.2& 10& 2.70& 9.26& 1&1 \\ \hline
CPLMFC ($\alpha=2$)\tnote{3}&2.116& 15.39& 1.21& 0.78& 0.55& 2.22&0.6& 0.25\\ \hline
CPLMFC ($\alpha=16$)\tnote{3}& 1.160& 9.27& 1.21& 6.24& 4.39& 17.74&0.6& 0.25 \\ \hline
CPLMFC ($\alpha=16$)\tnote{3}& 1.338& 18.15& 1.13& 6.24& 4.39& 17.74&0.25& 0.25 \\
\bottomrule
\end{tabular}
\begin{tablenotes}
\item[1] unit-step input load disturbance
\item[2] unit-step setpoint without load disturbance.
\item[3] unit-step setpoint with unit-step input load disturbance.
\end{tablenotes}
\end{threeparttable}
}
\end{table*}

\subsection{Delayed Integrating, Nonlinear and Uncertain System}
\begin{figure}[!t]
\centering
\includegraphics[width=3.5in]{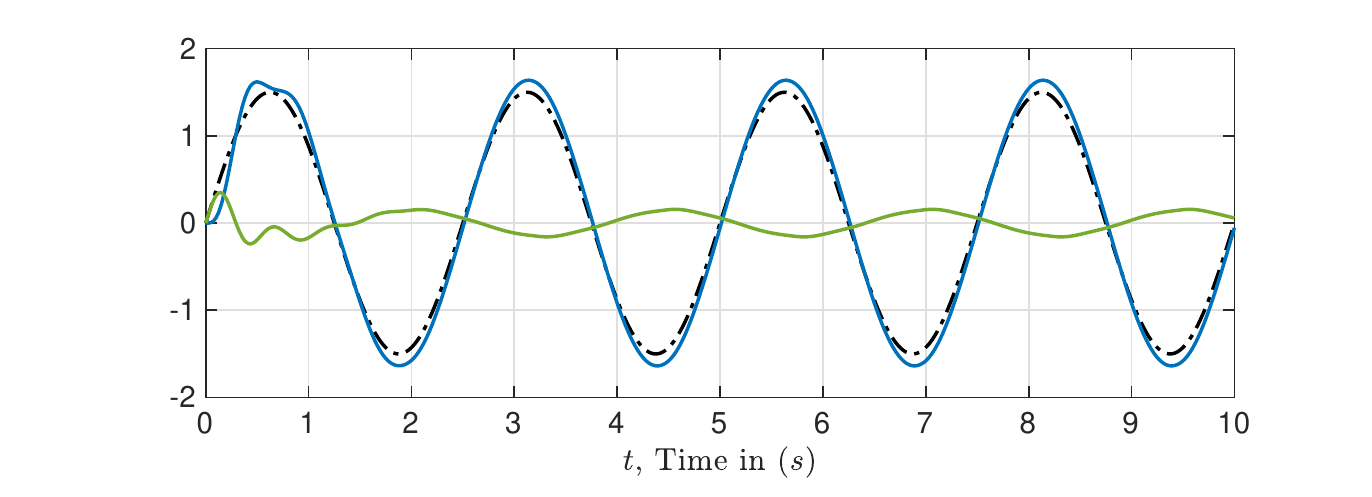}%
\caption{Non-minimum phase (double-integrating) scenario for $P_2(s)$ ($b=0$, $\tau_l=10\tau$, $m=5.4$): Desired Trajectory (dash-dotted, black), Position Response (solid blue), Tracking Error Response (solid green).}
\label{figpmlm}
\end{figure}
\begin{figure}[!t]
\centering
\includegraphics[width=3.5in]{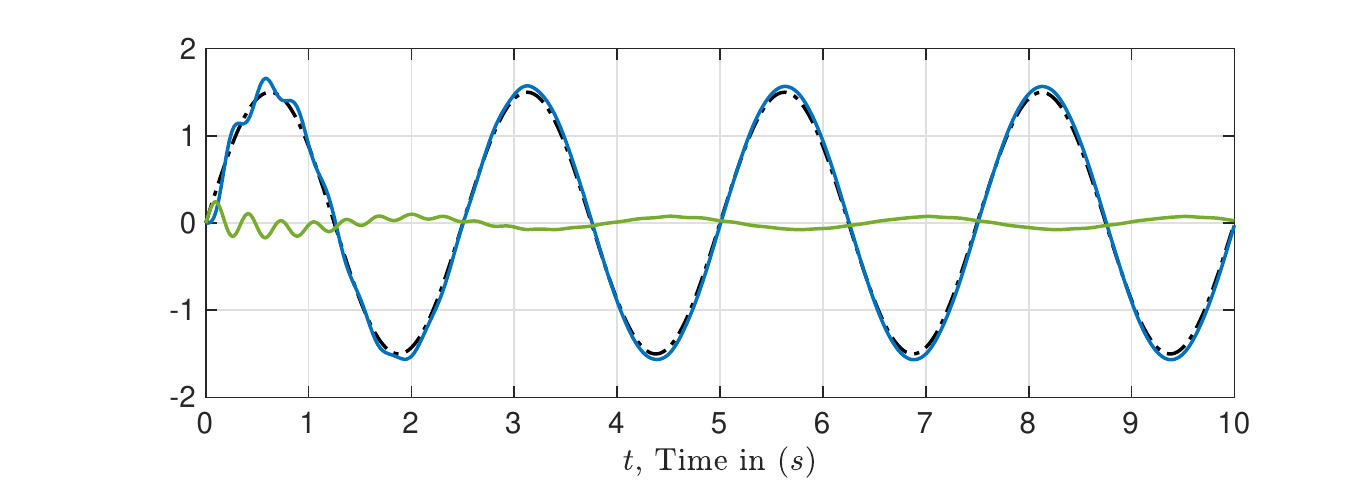}%
\caption{Non-minimum phase (single right-half plane pole) scenario for $P_2(s)$ ($b=-35.1$, $\tau_l=10\tau$, $m=5.4$): Desired Trajectory (dash-dotted, black), Position Response (solid blue), Tracking Error Response (solid green).}
\label{figpmlm2}
\end{figure}
One commonly used dynamical system for precise repetitive positioning applications is the permanent-magnet linear motor (PMLM) model. This system was used as a benchmark case in \cite{tanOnlineAutomaticTuning2007} and a similar model in \cite{zhangTheoryDesignPID2019a}. The PMLM can be modeled as an integrating, nonlinear and uncertain system with delay dynamics. The nonlinear model is expressed in a compact transfer-function form (\ref{pltpmlm}) relating position $y$ to input voltage $u$. The initial identified system parameters are: the mass constant $m=5.4$, the damping constant $b =35.1$, and the delay-time $\tau_l=0$. The aim is to track a sinusoidal reference trajectory, $r=A\sin\textstyle{( \frac{{2\pi}f\,k}{N} )}$, where the amplitude $A=1.5\,\mathrm{{}m}$, frequency $f=4\,\mathrm{Hz}$, while $k$ and $N$ correspond to the current and total discrete-time counts respectively. The frictional force $f_F$ and ripple force $f_R$ account for noise, disturbances and uncertainty in the system.
\begin{IEEEeqnarray}{c}\label{pltpmlm}
P_2\left( s\right) = \frac{y}{v}=\frac{1}{s\left( m\,s + b \right)}\\
v=\left(8.1\,u\right)e^{-\tau_l s} -f_F-f_R\\
\noalign{\noindent where\vspace{\jot}}
f_R={3}\sin\left( \frac{2\pi\,y}{0.0712} \right) \IEEEnonumber\\
f_F=\left( 3+10\left| \dot{y} \right| \right) \mathrm{sgn}\left( \dot{y} \right) \IEEEnonumber
\end{IEEEeqnarray}
The sampling-time was taken as $\textstyle{\tau=\frac{1}{1000}}$~seconds. Applying the identification procedure in section~\ref{Stsident}, the estimated settling-time for the system is $1.8$~seconds.
As illustrated in Table~\ref{tabpmlm}, here we use three cases to cover  future variations in the system parameters, namely: mass-variation $m\in\left[1, 5.4\right]$, damping variation $b\in\left[35.1, 0, -35.1 \right]$, and delay variation $\tau_L\in\left[0, 10\tau\right]$. This amounts to eight scenarios.
We proceed in Case~1 with the normal integrating scenario with mass and delay variations. Considered next in Case~2 is the double-integrating scenario with mass variations. Then in Case~3, the integrating scenario with a pole in the right-half $s$-plane and delay variations is considered.
\begin{table}[!t]
\centering
\begin{threeparttable}
\renewcommand{\arraystretch}{1.3}
\caption{System Parameters, Performance Measures and Controller Settings for Plant $P_2=\frac{1}{s\left( ms+b \right)}$.}
\label{tabpmlm}
\centering
\footnotesize{
\begin{tabular}{*{1}{c}*{3}{c}||*{2}{c}||*{3}{c}}
\toprule
\hline
Case&$m$ & $b$ & $\tau_l$ & $\bm{J_{iae}}$& $\bm{J_{ise}}$& $\alpha$& $\lambda_i$& $\lambda_d$ \\
1&$5.4$& $35.1$& $0$& ${0.68}$ &${0.06}$ &$500$ &$0.5$ &$0.01$\\ \hline
1&$1$& $35.1$& $0$&  ${0.65}$ &${0.05}$ &$500$ &$0.5$ &$0.01$\\ \hline
1&$5.4$& $35.1$& $10\tau$&  ${1.12}$ &${0.16}$ &$500$ &$0.5$ &$0.01$\\ \hline
1&$1$& $35.1$& $10\tau$&  ${1.09}$ &${0.15}$ &$500$ &$0.5$ &$0.01$\\ \hline
\hline
2&$5.4$& $0$& $10\tau$& ${1.02}$ &${0.13}$ &$250$ &$0.8$ &$0.1$\\ \hline
2&$1$& $0$& $10\tau$&  ${0.94}$ &${0.11}$ &$250$ &$0.8$ &$0.1$\\ \hline
\hline
3&$5.4$& $-35.1$& $0$& ${0.49}$ &${0.03}$ &$500$ &$0.8$ &$0.1$ \\ \hline
3&$5.4$& $-35.1$& $10\tau$& ${0.52}$ &${0.04}$ &$500$ &$0.8$ &$0.1$\\ \hline
\bottomrule
\end{tabular}
}
\end{threeparttable}
\end{table}

The tuning settings in Case~1 start with safe values of $\lambda_i$ and $\lambda_d$. The proportional gain is tuned with $\alpha$. Recall, that the stabilizing $k_p$ set is unknown. In this simulation, $\alpha$, starting from a smaller value is increased to $\alpha=500$. This value is not unique, a lower or higher value may be chosen. However, too high a value may excite the closed PID-loop to instability. In Case~2, since it is a more difficult scenario, the value of $\alpha$ is reduced by half. Also, the belief on the derivative is slightly increased by a factor of $10$. The belief on the integral action may be left unchanged, but here, it is increased to $0.8$ to reduce the tracking error. Further, in Case~3, $\alpha$ was increased back to $500$, as a lower or too high value may further destabilize the already unstable system. A more robust value applicable to all instances in Table~\ref{tabpmlm} is $\alpha=500, \lambda_i=0.5, \lambda_d=0.1$.

Figure~\ref{figpmlm} shows the response for a non-minimum phase double-integrating scenario where $b=0$, $\tau_l=10\tau$, and $m=5.4$. Also, Figure~\ref{figpmlm2} shows the response for a non-minimum phase worst-case scenario where $b=-35.1$, $\tau_l=10\tau$, and $m=5.4$. The results in Table~\ref{tabpmlm} and tracking error responses in Fig~\ref{figpmlm}--\ref{figpmlm2}, accentuate the promise of closed-PID loop performance through the CPLMFC method, with respect to the $J_{iae}$ and the $J_{ise}$ indices.

\section{Discussions}\label{secdiss}
It is straightforward to see that the only parameters to manually adjust are the hyper-parameter $\alpha$ and the critic weights $\lambda_i, \lambda_d$. If properly set, therefore abstracting the setting of the PID gains, they automatically determine the appropriate PID settings.

In terms of representation, the critic-PID control form in section~\ref{sec2}, although in simple form can be viewed as computational intelligence. It reflects the presence of criticism using critic weights to fault each of the three PID contributing terms. Using knobs, the critic weights may also represent the belief of a control-loop operator in order to give room for manual oversight. If the critic weights are set such that $\lambda_d \simeq 0, \lambda_i \neq 0$, the PID form reduces to PI-only control, while $\lambda_i \simeq 0, \lambda_d \neq 0$, will reduce the form to PD-only control.

However, since a limitation in this work, is that the critic gains were manually chosen. An important question going forward then is how $\lambda_i,\lambda_d$, the two adjustable critic weights of the PID output terms can be automatically set to improve PID control performance. This could be a future basis for fusing an internal concept of learning through critic weights into PID formulations. Therefore the design of a stable real-time learning or optimization framework for criticizing the three contributing PID terms will be a challenging but interesting problem for future research.

Also, results of the simulations have further shown that this CPLMFC approach to PID tuning can not only guarantee but also streamline the search for accurate and stable PID control gains for controlling dynamical systems. However, it is useful to note that this method cannot stabilize: One, systems that cannot approach a finite settling time behaviour; Two, systems that the CPLM poles cannot dominate in closed-loop.

For many dynamical systems that are not difficult to control with respect to stability and dead time, for example: electric motor drives, then regulating such systems around their operating regions can become less complex and more automatic by using an input-output approach to identifying their settling time behaviour in a closed PID-loop. We note that, we have successfully applied the CPLMFC method to control the speed of two electric dc-motors in a differential drive mobile robot. The whole process involved will be detailed in a future paper. One insight that the CPLMFC method has demonstrated again, is that PID control is highly dependent on the knowledge of time \cite{clairControllerTuningControl1993}. This can be referred to as time-fitting.

The settling-time identification procedure described in this paper is not automatic. Therefore, it will be of practical interest to fully automate and develop this aspect. Since automatic adaptive real-time parameter settings brings up the question of safety, consequently, there is a further need to fully investigate the safety guarantees of the CPLMFC approach.

Also, another limitation in this paper is that the adaptive rule for the determination of the proportional gain in the unknown stabilizing and optimal set, was augmented for delay-dynamics in the closed-loop system. Therefore, extension of the PMF theorems in section~\ref{Sadaptkp} to consider the presence of delay-dynamics will be a significant improvement.

\section{Conclusions}\label{secconc}
The CPLMFC results in this paper provide a method for automatic robust PID design in the model following sense based on the characteristic settling behaviour of stabilizable dynamical systems. The main idea is that if we can appropriately identify a stabilizable proportional gain for the closed PID-loop model (CPLM), and appropriate critic weights, then both stable and accurate settling behaviour can be guaranteed for the actual closed PID-loop system.

The main contributions in this paper have been highlighted in section~\ref{Smaincontrib}. Continuous improvement of the CPLMFC theory will be of interest. Also, for further investigations, the application of this method to a wide range of practical dynamical systems will be useful.




\bibliographystyle{unsrt}
\bibliography{../bib/MastersPID}

\appendices

\section{Type-1 FIS Mapping}\label{fisapp}
\begin{defn}\label{fisfbfdef}
For input $\bm{x} = \left[ b \quad c \right]^T$, the Type-1 FIS that maps to $\bm{y} =\left[ x_s \quad x_{pk} \right]^T$  is represented as the fuzzy basis function expansion:
\begin{IEEEeqnarray}{c}
\bm{y} \left( \bm{x} \right)=\sum_{l=1}^M{c_{0}^{l}\phi _{j}^{l}\left( \bm{x} \right)}\\\label{fisdef}
\phi _{j}^{l}\left( \bm{x} \right) =\frac{\prod_{i=1}^p{\mu _{F_i}^{l}\left( x_i \right)}}{\sum_{l=1}^M{\prod_{i=1}^p{\mu _{F_i}^{l}\left( x_i \right)}}}\\\label{fbfdef}
\mu _{F_i}^{l}\left( x_i \right) =\begin{cases}
	\mu _L\left( x \right) =\mathrm{nlsig}^-\left( x;\bar{c}_L,\bar{d}_L \right) ;x<\frac{\bar{c}_L+\bar{c}_R}{2}\\
	\mu _R\left( x \right) =\mathrm{nlsig}^+\left( x;\bar{c}_R,\bar{d}_R \right) ;x\geqslant \frac{\bar{c}_L+\bar{c}_R}{2}\\
\end{cases}\label{cnlsig}\IEEEnonumber
\end{IEEEeqnarray}
\end{defn}
For $\mathrm{nlsig}^-$, the following constrains hold $x_{\min}^-=\bar{c}_L-\bar{d}_L$, $x_{\max}^-=\bar{c}_L$, $\bar{c}_L>x_{\min}$ and for $\mathrm{nlsig}^+$, the constrains are
$x_{\min}^+=\bar{c}_R$, $x_{\max}^+=\bar{c}_R+\bar{d}_R$, $\bar{c}_R<x_{\max}$. Also, $y_{\max}=\text{1, }y_{\min}=0$.
\begin{table}
\centering
\renewcommand{\arraystretch}{1.3}
\footnotesize{
\caption{Type-I FIS Design Choice}
\label{fischoice}
\begin{tabular}{*{3}{c}}
\toprule
FIS Type& \multicolumn{2}{c}{Takagi-Sugeno-Kang (TSK)} \\ \hline
Input(s) & \multicolumn{2}{c}{$b$ and $c$ $\therefore p=2$} \\ \hline
Antecedent Fuzzifier & \multicolumn{2}{c}{Singletons} \\ \hline
Universe & \multicolumn{2}{c}{ $0-1$} \\ \hline
Output(s) & $x_{pk}$ & $x_{ts}$ \\ \hline
Consequent Fuzzifier & Singleton & Non-Singleton \\ \cline{2-3}
Universe & $2-6$ & $0-20$ \\ \cline{2-3}
T-norm  & Product & Product\\ \cline{2-3}
T-conorm & Nil & Max \\ \hline
MF (Parameterized) & \multicolumn{2}{c}{closed $n$-logistic sigmoids, see (\ref{cnlsig})}\\ \hline
MF Parameters & \multicolumn{2}{c}{Pre-specified, see Table~\ref{xfis}} \\ \hline
Number of Rules & \multicolumn{2}{c}{$M = 121$ for each output} \\
\bottomrule
\end{tabular}
}
\end{table}

\section{N-Logistic Sigmoid Function \cite{somefunSomefunAgbaNlogisticsigmoidNlogistic2020}}\label{nlsigapp}
\begin{defn}\label{nlsigdef}
The $\mathrm{n}$-logistic sigmoid function, where $\delta \in \mathbb{R}^{n\times 1}$ and $\kappa _x,\kappa _y\in \mathbb{R}^{\left( n+1 \right) \times 1}$ with $\lambda =\text{6}$ as a standard default value is defined as:
\begin{IEEEeqnarray}{c}
y=\mathrm{nlsig}^{\pm}\left( x;x_{\max},x_{\min},y_{\max},y_{\min},n,\lambda, \xi \right)\\\IEEEnonumber
=\kappa _{y,1}+\sum_{i=1}^n{\frac{\kappa _{y,i+1}-\kappa _{y,i}}{1+e^{\pm \alpha \left( x-\delta _i \right)}}}\\
\mbox{where:}\quad \xi =\begin{cases}
0; \quad\mathrm{nlsig}^-\\
1; \quad\mathrm{nlsig}^+\\
\end{cases}\\
\varDelta _x=\frac{x_{\max}-x_{\min}}{n}, \quad \varDelta _y=\frac{y_{\max}-y_{\min}}{n}\\
\alpha =\lambda \frac{2}{\kappa _{x,i+1}-\kappa _{x,i}}=\lambda \frac{2}{\kappa _{x,2}-\kappa _{x,1}}\\
\kappa _{x,i+1}=\kappa _{x,i}+\varDelta _x, \quad \kappa _{y,i+1}=\kappa _{y,i}+\varDelta _y\\
\delta _i=\frac{\kappa _{x,i+1}+\kappa _{x,i}}{2},\quad i=\text{1,...,}n
\end{IEEEeqnarray}
such that $\kappa _x=\left[ \kappa _{x,i},...,\kappa _{x,i+1} \right],\,\kappa _{x,1}=x_{\min},\,\kappa _{x,n+1}=x_{\max}$ and $\kappa _y=\left[ \kappa _{y,i},...,\kappa _{y,i+1} \right],\,\kappa _{y,1}=y_{\min},\,\kappa _{y,n+1}=y_{\max}$.
\end{defn}
The following conditions are satisfied:
$\lim_{x\rightarrow x_{\min}} \text{nlsig}^-\left( x \right) =y_{\min}$,
$\lim_{x\rightarrow x_{\max}} \text{nlsig}^-\left( x \right) =y_{\max}$,
$\lim_{x\rightarrow x_{\min}} \text{nlsig}^+\left( x \right) =y_{\max}$,
$\lim_{x\rightarrow x_{\max}} \text{nlsig}^+\left( x \right) =y_{\min}$

\end{document}